\documentclass[pre,twocolumn,showpacs,preprintnumbers,amsmath,amssymb,unsortedaddress]{revtex4-1}

\usepackage{amsmath}
\usepackage{amsfonts}
\usepackage{amssymb}
\usepackage{mathrsfs}
\usepackage{mathptmx}
\usepackage{graphicx}
\usepackage{epstopdf}
\usepackage{listings}
\usepackage{color}
\usepackage{times}
\usepackage{tabularx}

\newcommand{\inlinemaketitle}{{\let\newpage\relax\maketitle}}

\setcitestyle{super}
\makeatletter % Reference list option change
   \renewcommand\@biblabel[1]{#1.} % from [1] to 1.
\makeatother

\begin{document}

\title{Manipulation of Confined Polyelectrolyte Conformations Through Dielectric Mismatch}

\author{Trung Dac Nguyen$^1$}
\author{Monica Olvera de la Cruz$^{1,2}$}
\email{m-olvera@northwestern.edu}
\affiliation{$^1$Department of Chemical and Biological Engineering, Northwestern University, Evanston, IL 60208}
\affiliation{$^2$Department of Materials Science and Engineering, Northwestern University, Evanston, IL 60208}

\date{\today}

\begin{abstract}

ABSTRACT: We demonstrate that a highly charged polyelectrolyte confined in a spherical cavity undergoes reversible transformations between amorphous conformations to a four-fold symmetry morphology as a function of dielectric mismatch between the media inside and outside the cavity. Surface polarization due to dielectric mismatch exhibits an extra ``confinement'' effect, which are most pronounced within a certain range of the cavity radius and the electrostatic strength between the monomers and counterions and multivalent counterions. For cavities with a charged surface, surface polarization leads to an increased amount of counterions adsorbed in the outer side, further compressing the confined polyelectrolyte into a four-fold symmetry morphology. The equilibrium conformation of the chain is dependent upon several key factors including the relative permittivities of the media inside and outside the cavity, multivalent counterion concentration, cavity radius relative to the chain length, and interface charge density. Our findings offer insights into the effects of dielectric mismatch in packaging and delivery of polyelectrolytes across media with different relative permittivities. Moreover, the reversible transformation of the polyelectrolyte conformations in response to environmental permittivity allows for potential applications in biosensing and medical monitoring.


\end{abstract}
\maketitle

\label{sec:Intro}

The thermodynamic and structural behaviors of a highly charged polymer confined into a volume comparable to its size have been of interest both practically and fundamentally. In gene delivery applications,\cite{Fuller07,Kler12} for examples, DNA and RNA molecules are packaged inside protein capsids and lipid membranes. In living cells, proteins and DNAs are organized into micrometer-sized liquid droplets (also  known as membraneless organelles), whose interior resembles an organic solvent with a lower relative permittivity than the cytoplasm.\cite{Gomes18,Nott16} In such conditions, the difference in the relative permittivity between the media inside and outside the confinement space leads to polarization effects, which manifest themselves as induced charges at the interface between the two media. An insightful understanding of the collective effects of spatial confinement and dielectric mismatch on the polyelectrolyte conformational behavior is therefore crucial for efficient packaging, stabilization and transfer of polyelectrolytes across media with different dielectric constants. Furthermore, if the confined polyelectrolyte conformations can be manipulated through changes in the relative permittivity of the outside medium, it would open exciting possibilities for engineering stimuli-responsive nanodevices.

\begin{figure*}[ht!]
  \centering
  \includegraphics[width=0.75\textwidth, trim=0cm 0cm 0cm 0cm, clip=true]{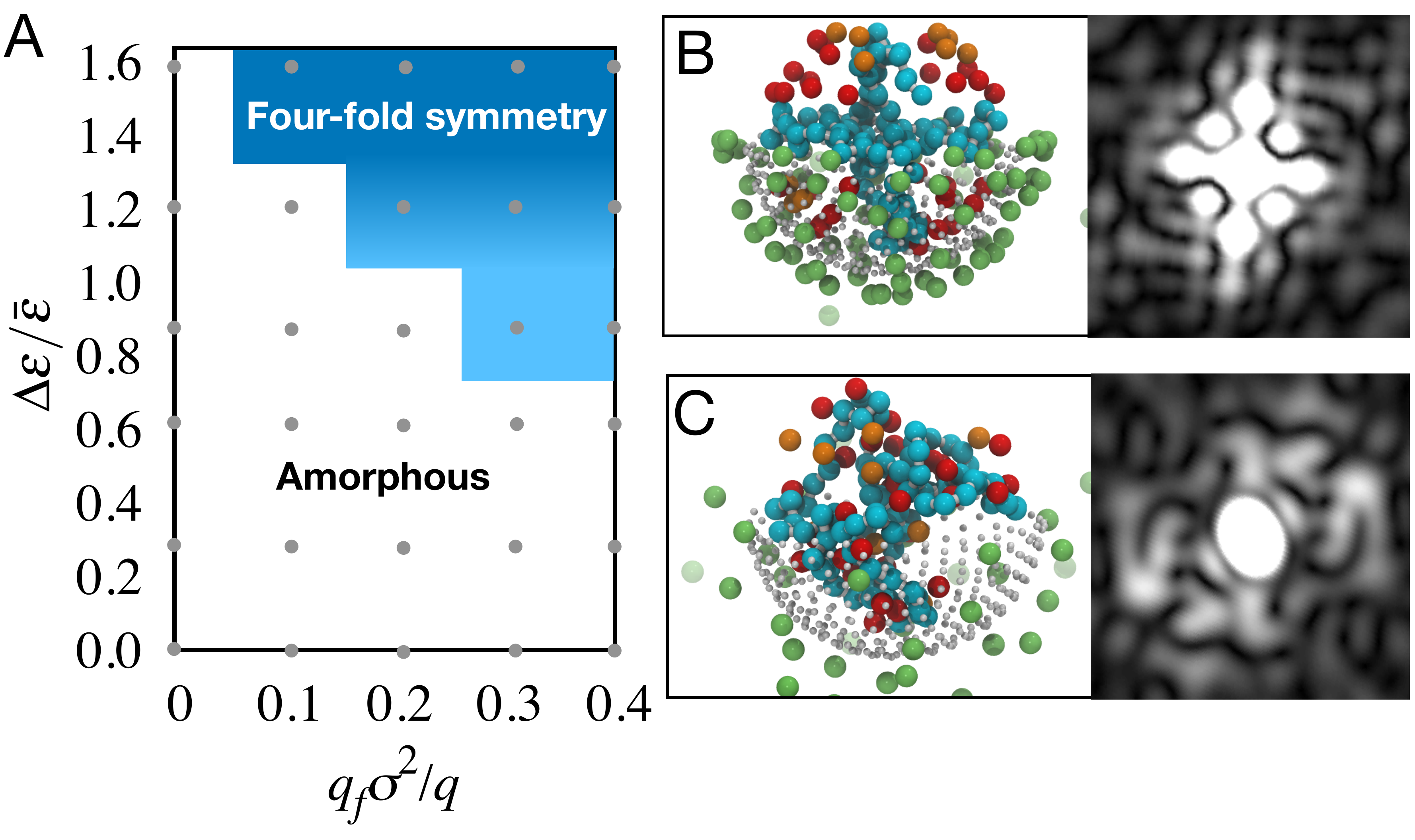}
  \caption{A) Phase diagram of equilibrium morphologies of the confined polyelectrolyte as function of the surface charge density ($q_f\sigma^2/q$) and dielectric mismatch ($\Delta \epsilon/\bar{\epsilon}$). Dots represent the state points where simulations are performed. Representative snapshots of B) the four-fold symmetry conformations, and C) amorphous conformations. Insets are the corresponding 2-D diffraction patterns of the monomers (\textit{i.e.}, the spheres in cyan). The diffraction patterns are taken along to the four-fold symmetry axis. The cavity radius is $R = 6\sigma$ and trivalent counterion fraction $\phi_m = 0.3$. Boundaries between the morphologies are sketched to guide the eye. \textcolor{black}{For B) and C), monomers are in cyan, monovalent counterions in orange, trivalent counterions in red, outer counterions in green; and small gray spheres are interface beads. For interpretation of the references in the text to color in this figure, the reader is referred to the web version of this article.}}
  \label{fig:phasediagram}
\end{figure*}

The influences of spatial confinement on the polyelectrolytes have been extensively investigated.\cite{Pais02,Angelescu06,Angelescu07,Hu08,Kumar08,Wang11,Giustini13,Nunes15,Shojaei17,Cordoba17,Cao17} The equilibrium conformation of a highly charged polymer chain confined in a cavity is evidently governed by the interplay between energetic and entropic factors coming from the confinement shape and size relative to the chain length, the chain bending stiffness, the electrostatic interaction between the charged monomers, counterions and the solvent, as well as their configurational entropy. For short flexible polyelectrolytes confined in a neutral spherical cavity, Kumar and Muthukumar \cite{Kumar08} using self-consistent field theory demonstrated that, for a given radius of the spherical cavity and fixed charge density along the backbone of the chain, solvent and small ion entropies dominate over all other contributions to the free energy. As the chain length increases, the contribution of chain conformational entropy and polymer-solvent interaction energy to the free energy become more pronounced. The effects of confinement on the chain conformation and counterion condensation have also been investigated.\cite{Pais02,Nunes15} Nunes \textit{et al.} \cite{Nunes15} found that the variation on the degree of condensation depends on counterion valence. For monovalent counterions, there is a minimum in the degree of ion condensation as a function of the confinement volume. Whereas, for trivalent ions, the degree of ion condensation always decreases as the confinement space is reduced. Notably, most of the particles reside close to the spherical wall and for the case where both the chain and counterions are confined, the counterion density profile peaks close to the wall.

\textcolor{black}{The nontrivial effects of surface polarization due to dielectric mismatch have also been addressed in numerous studies in the literature.\cite{Santos15,Santos16,Jing15,Anita18,Chu18,Shen17,Javidpour19,WuLi18}} It is shown that \textcolor{black}{polarization} effects become particularly pronounced at high dielectric contrasts,\cite{Santos15,Santos16,Jing15,Anita18,Chu18} in the presence of multivalent ions,\cite{Shen17,Javidpour19} and at highly curved interfaces.\cite{WuLi18,Javidpour19} Essentially, the electric field generated by the ions inside the confinement space polarizes the interface between the media, leading to nonzero induced charges that attract or repel the ions approaching the interface. de Santos and coworkers \cite{Santos16} studied the adsorption of polyelectrolytes on a flat like-charged surface and found that the monomer density profile is altered remarkably when the dielectric constant of the surface is much lower than the solvent. Shen \textit{et al.} demonstrated that the contribution of polarization to the net electrostatic interaction between two water droplets containing heavy metallic ions becomes significant when the droplets are submerged in oil.\cite{Shen17} Recently, Chu and coworkers studied that the effects of spatially varying dielectric permittivity on the phase behavior of salt-doped block copolymers. They found that the phase diagram is shifted to lower values of block incompatibility $\chi N$, and becomes slightly asymmetric with a broader range of the neutral block fraction for the lamellar phase.\cite{Chu18}

In the present study, we \textcolor{black}{demonstrate the ability} of a highly charged polyelectrolyte confined spherical droplet to adopt distinct conformations in response to changes in the environmental permittivity. We show that the collective effects of spatial confinement and surface polarization due to dielectric mismatch influence the chain conformational free energy landscape, increasing the work required to change the equilibrium chain conformation. The cavity size, electrostatic strength between the monomers and counterions and multivalent counterions, and dielectric contrast are shown to have substantial effects to the polyelectrolyte conformations. Furthermore, the polyelectrolyte motions at short length scales are facilitated by the condensed counterions. The findings therein offer insights into the effects of dielectric mismatch, which were often overlooked in previous studies on confined polyelectrolytes, and suggest possibilities to manipulate the confined polyelectrolyte morphologies through controlling the relative permittivity of the outside medium.

%The paper is organized as follows. The model and simulation method are present in Section \ref{sec:model}. The effects of several key factors on the conformational behavior of the polyelectrolyte is analyzed (Section \ref{sec:pmf})
%followed by spatial distribution of the charged monomers and counterions (Section \ref{sec:profiles}).

\section{Results and Discussion}
\label{sec:results}
\textcolor{black}{The confined charged polyelectrolyte system under investigation is represented by a coarse-grained model (see Section \ref{sec:model} for specific details). Here we give a brief introduction to the relevant input parameters to set the stage for the results that follow. The polymer chain is represented by the conventional bead-spring model \cite{Kremer90} composed of $N$ coarse-grained monomers.  There are $N^{+}_c$ monovalent counterions and $N^{3+}_c$ trivalent counterions to neutralize the total charge of the polymer: $N^{+}_c + 3N^{3+}_c = N$. The relative permittivities inside and outside the cavity are $\epsilon_{in}$ and $\epsilon_{out}$, respectively. The spherical cavity has radius of $R$, the surface of which is discretized into $M$ vertices arranged in a regular triangular mesh. The charge at each vertex is $q_f 4\pi R^2/M$. When $q_f$ is nonzero, additional counterions are added outside the cavity to ensure charge neutralize.  The polymer beads and counterions interact with each other and with the confinement wall \textit{via} the Weeks--Chandler--Andersen (WCA) potential with the characteristic length, $\sigma$, and energy scale, $\epsilon$.\cite{WCA} The multivalent counterion fraction is defined as $\phi_m = N^{3+}/N$.}

\subsection{Polyelectrolyte conformational behavior}
\label{sec:qsurf}

\textcolor{black}{Within the linear electrostatics framework, polarization effects due to dielectric mismatch between the medium inside and outside the cavity essentially give rise to induced charges at the interface, which attract charged particles toward the medium with higher relative permittivity, and repel them away from the medium with lower relative permittivity.\cite{Electrodynamics}} We show in Fig. \ref{fig:phasediagram} the ``phase diagram'' that summarizes the equilibrium conformations (or morphologies) of the confined polyelectrolyte as function of dimensionless surface charge density ($q_f\sigma^2/q$) and dielectric mismatch ($\Delta \epsilon/\bar{\epsilon}$), where $\Delta \epsilon = \epsilon_{in} - \epsilon_{out}$ and $\bar{\epsilon} = (\epsilon_{in} + \epsilon_{out})/2$. Here the equilibrium conformations are defined as those correspond to the free energy minimum of the chain radius of gyration obtained from umbrella sampling simulations.

For positively charged interfaces ($q_f > 0$), when dielectric mismatch is sufficiently high, we observe an increase in the number of outer counterions (green spheres) attracted towards the interface, leading to a stronger compression to the polyelectrolyte inside the cavity. As a result, the polyelectrolyte adopts a morphology with a four-fold symmetry, as can be seen from the \textcolor{black}{bright spots} in the diffraction pattern of the monomers in Fig. \ref{fig:phasediagram}B \textcolor{black}{(also Supporting Information, Movie S1). The monovalent and trivalent counterions are attracted to the interface and divided into the four quarters created by the polyelectrolyte.}

When the dielectric mismatch is decreased \textcolor{black}{by increasing $\epsilon_{out}$ and keeping $\epsilon_{in} = 40$ fixed}, the amount of negatively charged counterions adsorbed to the interface from outside is substantially reduced. The net compression exerted on the polyelectrolyte by the adsorbed outer counterions is insufficient to induce the formation of the four-fold symmetry morphology. The packing of the monomers in the collapsed polyelectrolyte is thus amorphous, as indicated by the diffraction pattern in Fig. \ref{fig:phasediagram}C. The amorphous conformation is also observed for neutral interfaces ($q_f = 0$).

Importantly, we observe that the amorphous conformations are spontaneously transformed into the four-fold symmetry morphology when the polarization effects are turned on, and \textit{vice versa} (see Supporting Information, Fig. \ref{fig:transforms}). The transition between the two morphologies is rather a weak first-order transition, as indicated by \textcolor{black}{the increased fluctuation of the radius of gyration as a function of $\Delta \epsilon/\bar{\epsilon}$ and by the bimodal distribution of the radius of gyration within the transition within the transition range} (Supporting Information, Fig. \ref{fig:Rgvariance}).

\begin{figure*}[ht!]
  \centering
  \includegraphics[width=0.75\textwidth, trim=0cm 0cm 0cm 0cm, clip=true]{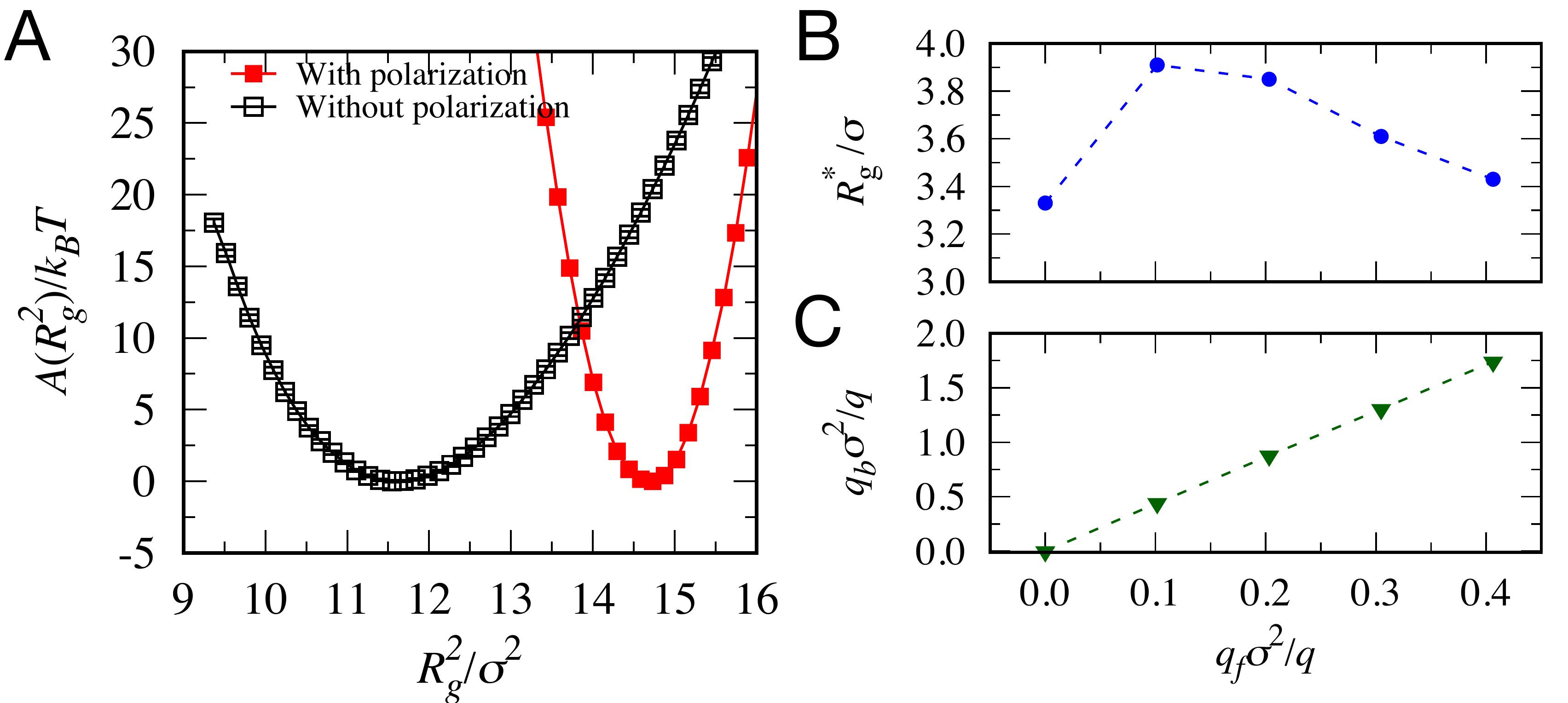}
  \caption{A) Free energy profile of the polyelectrolyte radius of gyration for surface charge density $q_f\sigma^2/q = 0.2$. The location of free energy minimum gives the equilibrium radius of gyration $R^*_g$. ``With polarization'' corresponds to $\epsilon_{in} = 40$ and $\epsilon_{out} = 4$; ``Without polarization'' corresponds to $\epsilon_{in} = \epsilon_{out} = 40$, or when polarization effects are excluded. B) Equilibrium radius of gyration, $R^*_g$, as a function of surface charge density, $q_f$. C) Surface induced charge density, $q_b$, at $R^*_g$ as a function of $q_f$. The cavity radius is $R = 6\sigma$. Dash lines are drawn to guide the eye. The trivalent counterion fraction is $\phi_m = 0.3$.}
  \label{fig:pmfqsurf}
\end{figure*}
%\begin{figure}[ht!]
%  \centering
%  \includegraphics[width=0.5\textwidth, trim=0cm 0cm 0cm 0cm, clip=true]{four-fold.pdf}
%  \caption{Effects of surface polarization with nonzero surface charge density ($q_f\sigma^2/q = 0.2$) on the equilibrium conformation of the polyelectrolyte A) with polarization, and B) without polarization. Insets are the corresponding 2-D diffraction patterns of the monomers. The cavity radius is $R = 6\sigma$ and trivalent counterion fraction $\phi_m = 0.3$.}
%  \label{fig:fourfold}
%\end{figure}

The polyelectrolyte robustly adopts the four-fold symmetry morphology in different cavity sizes, $R = 5\sigma$ and $7\sigma$ (Supporting Information, Fig. \ref{fig:radii}), \textcolor{black}{for other chain lengths, $N = 60$ and 200 (Supporting Information, Fig. \ref{fig:chainlengths}) and for different multivalent counterion fractions $\phi_m$ from 0 to 0.3 (Supporting Information, Fig. \ref{fig:ftri})}. Our supplementary simulation further shows that when all the bonds between the monomers are removed, \textcolor{black}{the monomers also arrange into a four-fold symmetry configuration (Supporting Information, Fig. \label{fig:nobonds}), indicating that such a configuration is not governed by the (linear) connectivity of the monomers. In this case, the inside counterions are divided into eight groups with a simple cubic symmetry, presumably to minimize the net repulsion between the groups.} \textcolor{black}{The phase diagram for a smaller multivalent counterion fraction ($\phi_m = 0.1$) is shown in Supporting Information, Fig. \ref{fig:phasediagram01}, showing that when $\phi_m$ is decreased, the transition from the amorphous to four-fold conformations shifts to higher values of surface charge density and dielectric mismatch.} 

\textcolor{black}{It is interesting to note that the four-fold symmetry conformations of the confined polyelectrolyte are reminiscent of one of the morphologies of symmetric diblock copolymers formed in spherical nanopores.\cite{Yu07} Yu and coworkers\cite{Yu07} using simulated annealing Monte Carlo calculations predicted various symmetry-breaking morphologies of the copolymers depending on the confinement degree and the adsorbing preference of the confining surface to one of the blocks. They found that when the confining surface is sufficiently repulsive to one of the blocks (type B) while attractive to the other (type A), the B-rich domain would have two, three, four or six struts, which are filled by the type A blocks. The number of the struts, hence the symmetry of the morphologies, depends on the confinement degree, defined as the ratio between the pore diameter, $D$, and the characteristic length of the assembled morphologies in bulk, $L_0$. In the present study, the inside counterions and charged monomers play the roles of type A and type B particles, respectively, while the outer counterions condensed at the interface due to dielectric mismatch set the surface preference by attracting the inside counterions and repelling the charged monomers.}

\textcolor{black}{In light of the work by Yu and coworkers,\cite{Yu07} we hypothesize that the formation of the four-fold symmetry conformations of the confined polyelectrolyte with the counterions filled in between results to a delicate interplay between 1) the spatial confinement degree ($N/R$), 2) dielectric mismatch ($\Delta \epsilon/\bar{\epsilon}$) with sufficiently weaker Coulombic interaction inside the cavity than that outside ($\epsilon_{in} > \epsilon_{out}$), and 3) the presence of the inside counterions that are attracted to the interface and divided into similar-sized groups, presumably to minimize the Coulombic repulsion between the groups. We have performed additional simulations to elucidate the roles of these factors. First, when the chain length is sufficiently small (\textit{i.e.}, $N = 20$) relative to the cavity radius ($R = 6\sigma$), or in other words, confinement effects become negligible (Supporting Information, Fig. \ref{fig:chainlengths2}A). Second, when the chain length is too large (\textit{i.e.}, $N = 250$), confinement effects are so strong that the excluded volume interaction between the monomers dominates and the quarters of the four-fold symmetry morphology are filled up (Supporting Information, Fig. \ref{fig:chainlengths2}C). Third, if the monovalent and trivalent counterions are located outside the cavity, leaving only the polyelectrolyte inside the cavity, the polyelectrolyte adopts a spool-like conformation (Supporting Information, Fig. \ref{fig:amorphous}A).\cite{Nunes15} Finally, when electrostatic interaction strength inside the cavity is sufficiently strong, \textit{e.g.} with a lower value of $\epsilon_{in} = 10$, the polyelectrolyte and the inside counterions always form globular amorphous conformations (Supporting Information, Fig. \ref{fig:amorphous}B). In this case, surface preference has little influence on the collapse of the polyelectrolyte and counterions.}

To further investigate the effects of dielectric mismatch on the polyelectrolyte conformational behavior, we compute the free energy profiles of the polyelectrolyte squared radius of gyration, $R^2_g$. Fig. \ref{fig:pmfqsurf}A reveals the major differences in the free energy profile between with and without polarization effects due to dielectric mismatch. (The latter case is equivalent to the limit of $\Delta \epsilon/\bar{\epsilon} = 0$.) In either cases the free energy landscape is fairly simple, \textit{i.e.}, with a single global minimum, corresponding to the collapsed state with an equilibrium size, $R^*_g$. However, when surface polarization effects due to dielectric mismatch are pronounced, the energy minimum valley becomes much narrower, and $R^*_g$ is shifted to a remarkably higher value. It is evident that the work required to compress and to stretch the polyelectrolyte out of the equilibrium size increases remarkably in the presence of polarization effects.

\textcolor{black}{The jump in $R^*_g$ when surface charge density $q_f$ change from zero to nonzero values (Fig. \ref{fig:pmfqsurf}B) corresponds to the change from amorphous to the four-fold symmetry conformations. Compared to the amorphous collapsed conformations, the four-fold symmetry conformations have larger $R^*_g$ because the charged monomers spread out into the two perpendicular planes. For $q_f > 0$, however, the equilibrium radius of gyration $R^*_g$ decreases monotonously with the increasing surface charge density $q_f$ (Fig. \ref{fig:pmfqsurf}B and Supporting Information, Figs. \ref{fig:pmfqsurfs} and \ref{fig:qsurfs}).} This is because the polyelectrolyte in the four-fold morphology is compressed further towards the cavity center. This trend is only valid when $q_f$ is sufficiently low so to keep the confined polyelectrolyte within the conformational entropy-dominated regime, as already pointed out by Wang and Muthukumar.\cite{Wang11} Since the four-fold symmetry conformations are not spherical, the radius of gyration $R^*_g$ is certainly not the ideal order parameter, but can serve as a reasonable metric for comparing the size of the morphologies obtained from different nonzero surface charge densities.

\textcolor{black}{The changes in the slopes of the free energy landscape and in the equilibrium radius of gyration} with respect to $q_f$ can be explained as follows. When dielectric mismatch is sufficiently large, the amount of the outer counterions that are drawn toward, and adsorbed on, the interface from outside is increased. Consequently, the surface induced charge on the interface is amplified in the presence of polarization effects (Fig. \ref{fig:pmfqsurf}C). The competition between the repulsion from the increased number of adsorbed outer negatively charged counterions and the attraction from the positively charged interface gives rise to the four-fold symmetry morphology of the polyelectrolyte.

\subsection{Conformational free energy profiles as a function of \textcolor{black}{confinement volume, multivalent counterion fraction and electrostatic strength inside the cavity}} 
\label{sec:params}
In this section we focus on the influences of several key parameters on the free energy profile of the chain squared radius of gyration: confinement volume, multivalent counterion fraction, and electrostatic strength inside the cavity, with surface charge density assumed to be zero ($q_f = 0$). In this case, surface polarization effects due to dielectric mismatch alone simply shift $R^*_g$ to a slightly smaller value (Supporting Information, Fig. \ref{fig:neutral}).

\subsubsection{Confinement volume}
\label{sec:volume}
\begin{figure}[ht!]
  \centering
  \includegraphics[width=0.45\textwidth, trim=0cm 0cm 0cm 0cm, clip=true]{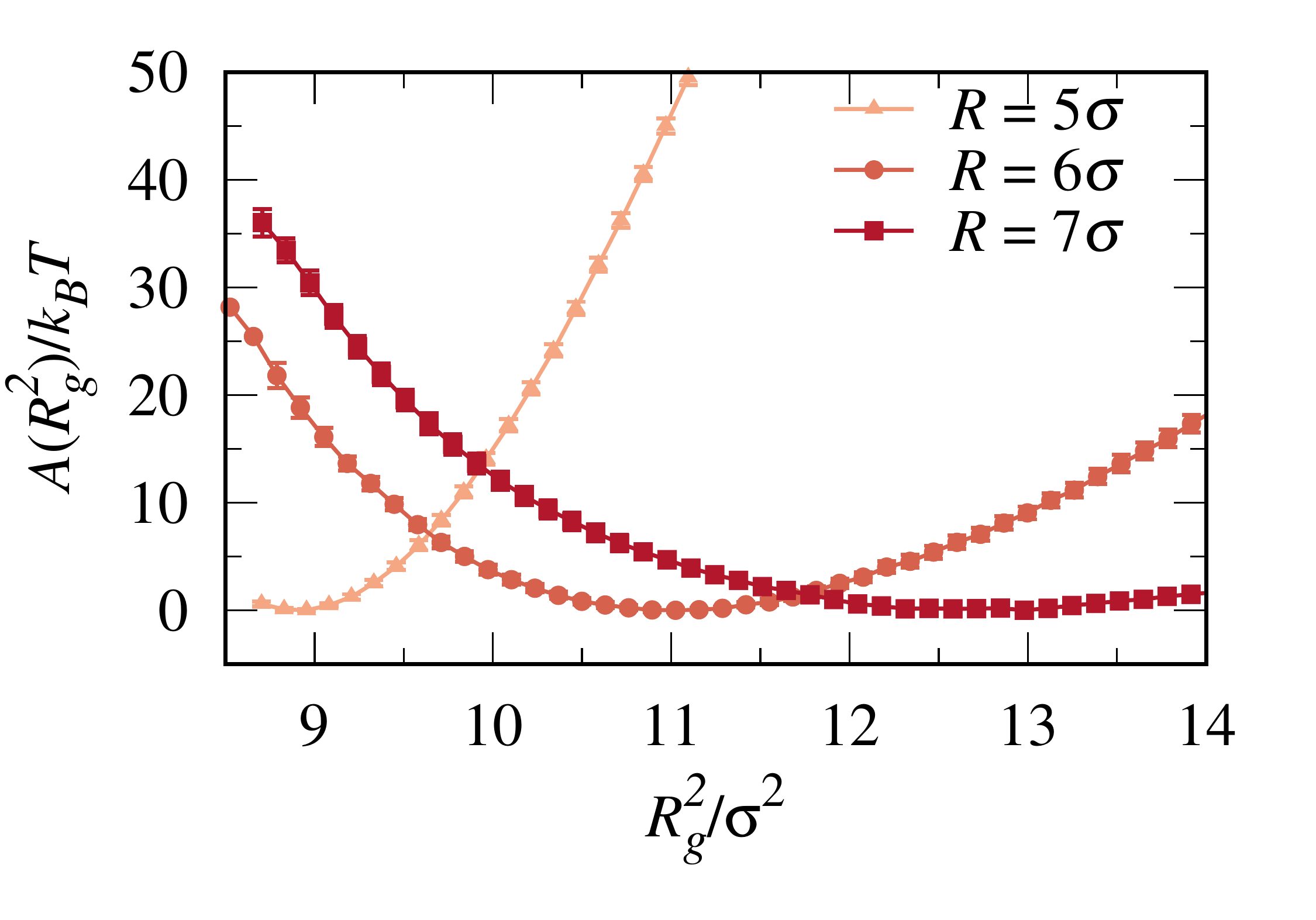}
  \caption{Effects of confinement size, $R$, on the free energy profile in the presence of surface polarization for $\phi_m = 0.3$. The relative permittivity outside the cavity is kept at $\epsilon_{out} = 4$. Interface charge density is $q_f = 0$.}
  \label{fig:confinement}
\end{figure}

To examine the effects of spatial confinement on the chain equilibrium conformations, we vary the cavity radius, $R$, keeping the chain length, $N$, unchanged. As can be seen from Fig. \ref{fig:confinement}, for a cavity radius as small as $R = 5\sigma$, the free energy almost monotonously increases with $R_g$, indicating that the chain is highly compressed by the spherical interface. This is the regime where spatial confinement is dominant and completely shadows surface polarization. As $R$ increases, the free energy minimum $R^*_g$ shifts to greater values, and the effects of spatial confinement is negligible for $R = 7\sigma$. It is interesting to note the remarkable change in the free energy profile when $R$ is increased from $5\sigma$ to $6\sigma$, with the equilibrium size $R^*_g$ expanded along with the reduced free energy gradient for $R_g > R^*_g$. \textcolor{black}{Likewise, when the chain length is increased for the same cavity radius, $R^*_g/N\sigma^2$ shifts to smaller values when the confinement degree increases (Supporting Information, Fig. \ref{fig:pmfchainlengths}.} Because the cavity size $R = 6\sigma$ indicates the regime where spatial confinement effects come in play, yet not too strong, we will focus on this value of the cavity radius in the following sections.

\subsubsection{Multivalent counterion fraction}
\label{sec:fraction}
The conformational features of polyelectrolytes in salt-free solution \cite{Stevens93,Stevens95,Gonzales95} and in the presence of monovalent and multivalent counterions \cite{Olvera95,Raspaud98,Solis00,Solis01,Frutos05,Henle05,Hsiao06,Sayar10,Sung16} have been thoroughly investigated. The precipitation of polyelectrolytes and the random coil-collapse transition of polyelectrolytes in a dilute solution induced by monovalent and multivalent counterions can be explained by thermodynamic models \cite{Olvera95,Raspaud98,Solis00,Solis01,Frutos05} and computer simulation.\cite{Pais02,Henle05,Hsiao06,Sayar10} Gonz\'{a}lez-Mozuelos and Olvera de la Cruz \cite{Gonzales95} developed a thermodynamic model using mean field approximation and short-range correlation corrections to investigate the equilibrium conformations of highly charged polyelectrolytes in salt-free dilution solutions. They demonstrated that the ratio between the counterion and monomer valencies is among the key factors that determine whether the collapsed or the rod-like conformations have lower free energy. \textcolor{black}{Frutos \textit{et al.} demonstrated that the tetravalent cation spermine, as a DNA condensing agent, reduces the pressure inside the capsid and influences the DNA ejection from the bateriophage.\cite{Frutos05}} It is evident that, at sufficiently high counterion concentration, the Coulombic attraction between the counterions and charged monomers is responsible for the random coil-collapse transition of the polyelectrolyte. The size of the collapsed chain is determined by the chain bulkiness and the counterion valency and concentration. This picture is indeed corroborated in our results.

\begin{figure}[ht!]
  \centering
  \includegraphics[width=0.45\textwidth, trim=0cm 0cm 0cm 0cm, clip=true]{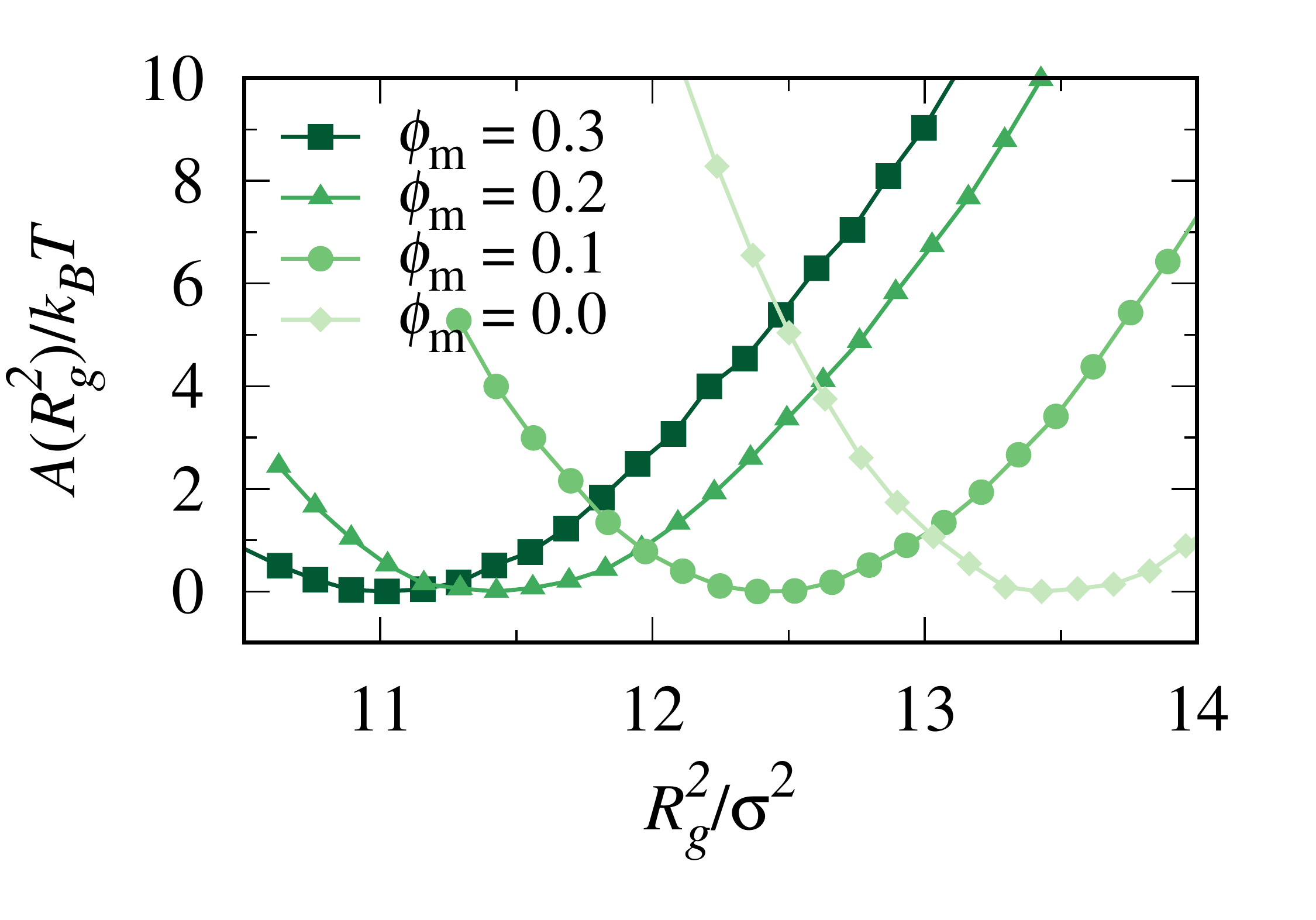}
  \caption{Effects of multivalent counterion fraction, $\phi_m$, on the free energy profile. The cavity radius is $R = 6\sigma$. Interface charge density is $q_f = 0$.}
  \label{fig:multivalent}
\end{figure}

The influences of the multivalent counterion fraction, $\phi_m$, on the conformational behavior of the polyelectrolyte are shown in Fig. \ref{fig:multivalent}. While the change in the slopes of the free energy profile is rather small, the collapsed radius $R^*_g$ decreases monotonously from $3.7\sigma$ to $3.3\sigma$ as $\phi_m$ increases from 0 to 0.3. This is because all the multivalent counterions are condensed on the polyelectrolyte and bridge different chain segments (Supporting Information, Fig. \ref{fig:probability}), and because trivalent counterions experience stronger repulsion from the interface due to surface polarization than monovalent counterions do. Whereas, the fraction of monovalent counterions condensed on the polyelectrolyte decreases as $\phi_m$ increases. The system minimizes its potential energy with trivalent counterions-monomer Coulombic attractions, and simultaneously gains additional translational entropy of the released monovalent counterions.

\begin{figure}[ht!]
  \centering
  \includegraphics[width=0.45\textwidth, trim=0cm 0cm 0cm 0cm, clip=true]{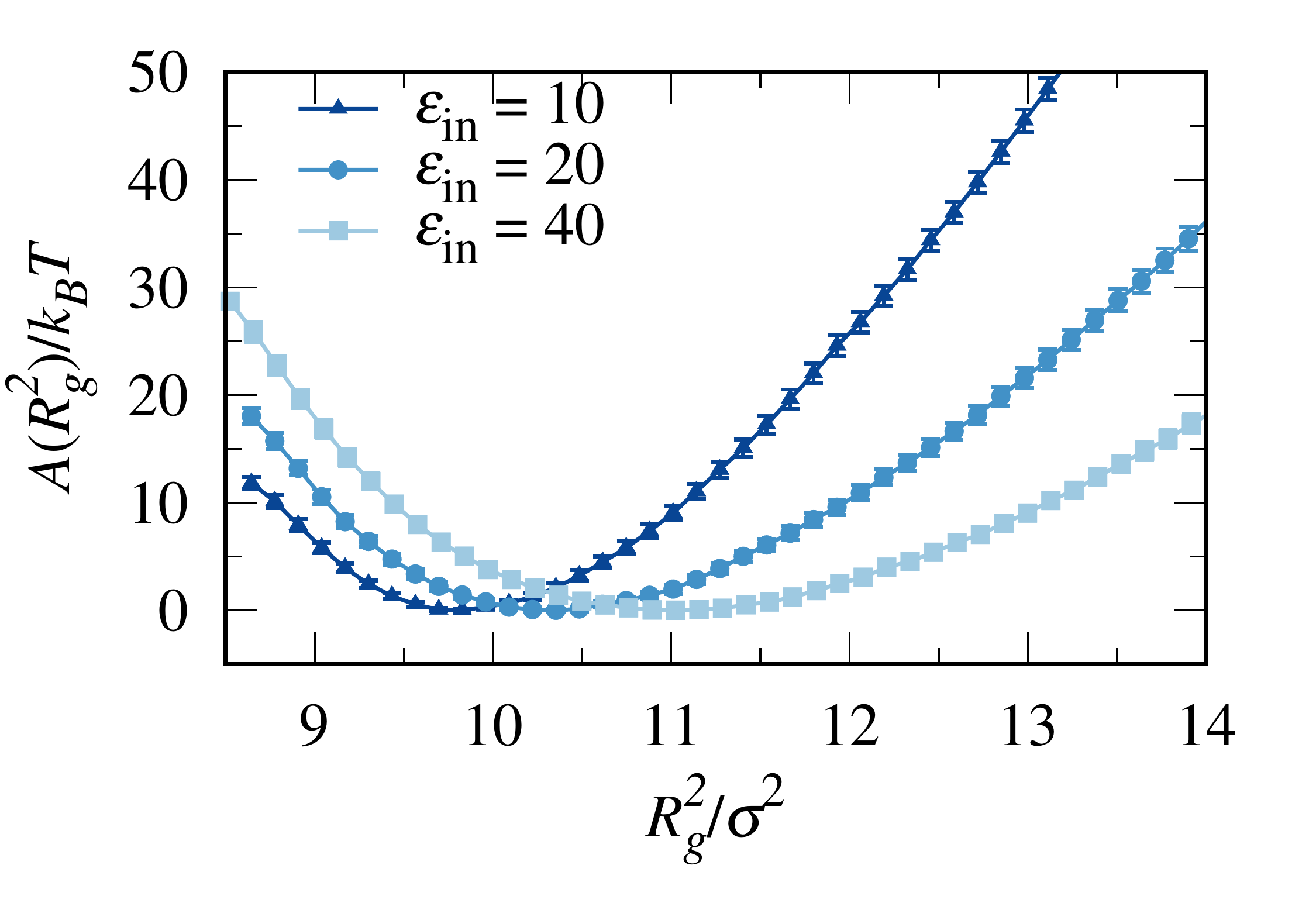}
  \caption{Effects of the electrostatic strength inside the cavity on the free energy profile in the presence of surface polarization for $\phi_m = 0.3$. The relative permittivity outside the cavity is kept at $\epsilon_{out} = 4$. Interface charge density is $q_f = 0$.}
  \label{fig:ein}
\end{figure}

\subsubsection{Electrostatic strength inside the cavity}
\label{sec:coulinside}
We show in Fig. \ref{fig:ein} the effects of the Bjerrum length inside the cavity $l_B \sim 1/\epsilon_{in}$ on the free energy profile by varying $\epsilon_{in}$. As $\epsilon_{in}$ decreases, $l_B$, and hence the strength of the Coulombic interaction between the charged particles, increases. On the one hand, the amount of counterions condensed on the polyelectrolyte increases. On the other hand, the repulsive forces between like-charged particles also become stronger. The combined effects of the Coulombic attraction and surface polarization, which pushes the charged particles away from the interface, then lead to the remarkable increase in the work needed to unravel the chain as $\epsilon_{in}$ decreases, hence the increased gradient of the free energy profile for $R_g > R^*_g$. Whereas for $R_g < R^*_g$, decreasing $\epsilon_{in}$ lowers the polyelectrolyte free energy due to the energetic gain associated with the increased Coulombic attraction between the monomers and counterions. These behaviors are in accord with the modified ``ion-bridging'' model developed for highly charged polyelectrolytes at high concentrations \cite{Olvera95,Raspaud98} and with the regime where the counterion concentration inside the cavity is sufficiently low to favor the collapse of polyelectrolyte.\cite{Hsiao06}

\subsection{Relaxation time of the Rouse modes of the polyelectrolyte}
\label{sec:rouse}
The influences of the counterions on the polyelectrolyte dynamics are certainly nontrivial, at the length scales of the monomers and of the sub-chains.\cite{Ting15,Webb18} Webb and coworkers showed that the dynamics of a single negatively charged polyelectrolyte is substantially suppressed across multiple length scales for different concentrations of lithium cations compared to that of neat polymers.\cite{Webb18} Here we also characterize the dynamics of the confined polyelectrolyte at different length scales through the relaxation time of its Rouse modes. The Rouse modes of a polyelectrolyte chain of length $N$ are defined as $\mathbf{X}_p(t) = (2/N)^{1/2} \sum^N_{i=1} \mathbf{r}_i(t) \cos[(p/N)(i-1/2)]$, where $p = 0, 1, 2, ... , N-1$.\cite{Kalathi15} The smallest mode ($p = 0$) captures the motion of the chain center of mass, whereas the other modes describe the motion of the segments composed of $(N-1)/p$ monomers.

\begin{figure}[ht!]
  \centering
  \includegraphics[width=0.5\textwidth, trim=0cm 0cm 0cm 0cm, clip=true]{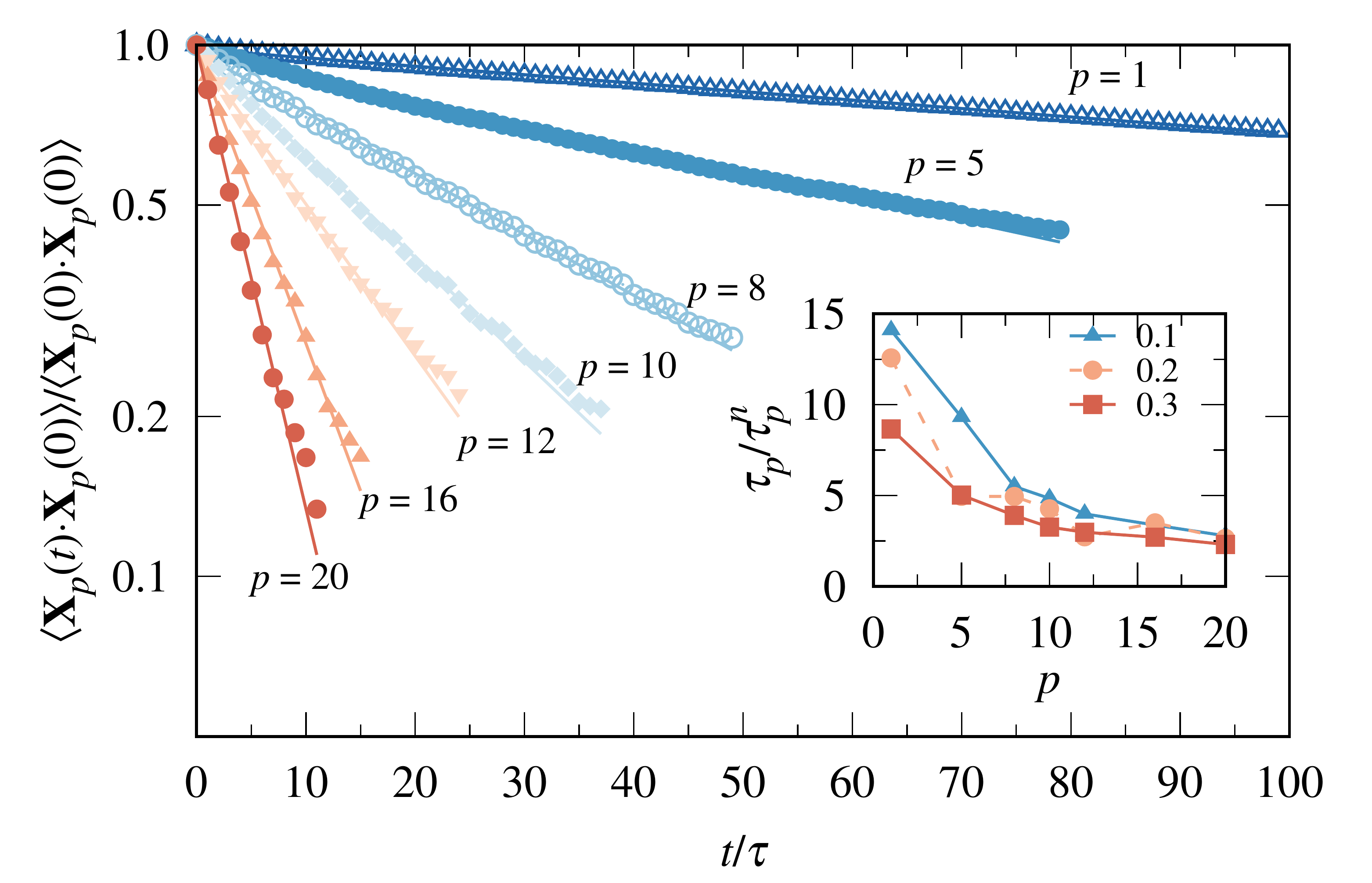}
  \caption{Normalized autocorrelation function of the Rouse modes of the confined polyelectrolyte in a neutral cavity ($q_f = 0$). The trivalent counterion fraction is $\phi_m = 0.3$. Inset shows the relaxation time for several Rouse modes, $\tau_p$, \textcolor{black}{normalized by the corresponding relaxation time of a neutral chain of the same length $N$, $\tau^n_p$, for different multivalent counterion fractions, $\phi_m$.}. The cavity radius is $R = 6\sigma$.}
  \label{fig:rouse}
\end{figure}

We compute the normalized autocorrelation function of the Rouse modes $\langle \mathbf{X}_p(t) \cdot \mathbf{X}_p(0) \rangle/\langle \mathbf{X}_p(0) \cdot \mathbf{X}_p(0) \rangle$, where $\langle \cdot \rangle$ is the average over multiple time origins. Fig. \ref{fig:rouse} shows the log plot of the autocorrelation function of several representative Rouse modes for $p \le 20$, \textcolor{black}{corresponding to the sufficiently long sub-chains to be consistent with the Rouse model. Up to a certain time interval, the autocorrelation function fits well with the simple exponentially decaying function  $e^{-t/\tau_p}$, where $\tau_p$ is defined as the relaxation time of mode $p$. $\tau_p$ decreases as $p$ increases, as indicated by the steeper slopes in the log plot.}  

\textcolor{black}{To evaluate the influences of the Coulombic interaction between the charged monomers and counterions on the dynamics of the polyelectrolyte, we compute the relaxation time of the Rouse modes of the chain with the Coulombic interaction switched off, that is, the non-bonded interaction between the monomers and counterions is only the WCA potential. As seen in the inset of Fig. \ref{fig:rouse}, $\tau_p/\tau^n_p > 1$ for all $p$, meaning that the relaxation times of the Rouse modes of the polyelectrolyte are always longer than those of a neutral chain. Particularly, the relaxation time at the length scale of the whole chain ($p = 1$) is greater than that of the neutral counterpart by an order of magnitude for all the values of $\phi_m$ studied. As $p$ increases, the relaxation time of the sub-chains at shorter length scales approaches, but is still longer than, that of the neutral chain. The longer correlation time in the dynamics of the polyelectrolyte across multiple length scales is attributed to the ionic correlation between the charged monomers and the counterions, which in consistent with the findings reported in the recent work by Webb and coworkers on ion-doped polymers.\cite{Webb18}} We note on passing that polarization effects alone have negligible effects on the relaxation time of the Rouse modes, at least for this particular case.

%For an ideal unperturbed chain, the relaxation time of mode $p$ is given by $1/\tau^0_p \sim \sin^2(p \pi/2N)$.

%\begin{table}[ht]
%\centering
%\begin{tabularx}{0.5\textwidth}{@{\extracolsep{\fill}} c c c c }
%  \hline
%  $p$ & $\phi_m = 0.1$ & $\phi_m = 0.2$ & $\phi_m = 0.3$ \\
%  \hline
%  1  & 0.00225 & 0.00252 & 0.00366 \\
%  5  & 0.00556 & 0.01044 & 0.01030 \\
%  8  & 0.01875 & 0.02085 & 0.02639 \\
%  10 & 0.03022 & 0.03431 & 0.04490 \\
%  12 & 0.04914 & 0.07193 & 0.06582 \\
%  16 & 0.10236 & 0.09884 & 0.12764 \\
%  20 & 0.16539 & 0.17426 & 0.19975 \\
%  \hline
%\end{tabularx}\caption{Inverse relaxation time of several Rouse modes, $1/\tau_p$ for different multivalent counterion fractions.}
%\label{tb:taup}
%\end{table}

\section{Conclusions}
\label{sec:conclusion}
We have demonstrated that a highly charged polyelectrolyte confined in a spherical cavity adopt two distinct morphologies, that is, the amorphous and four-fold symmetry conformations, depending on dielectric mismatch between the media inside and outside the cavity and surface charge density. The transition between the two morphologies is reversible upon changes in the relative permittivity of the outside medium. Surface polarization due to dielectric mismatch exhibits an extra ``confinement'' effect, which are most pronounced within a certain range of the cavity radius and the electrostatic strength between the monomers and counterions and multivalent counterions. Interestingly, for cavities with a charged surface, surface polarization leads to an increased amount of counterions adsorbed in the outer side, further compressing the confined polyelectrolyte into the four-fold symmetry conformations. The equilibrium conformation of the chain is dependent upon several key factors including multivalent counterion concentration, cavity radius relative to the chain length, and the relative permittivity of the medium inside the cavity. Our findings offer insights into the effects of dielectric mismatch in packaging and delivery of polyelectrolytes across media with different relative permittivities. Furthermore, it suggests that dielectric mismatch can be realized as a control knob to manipulate the confined polyelectrolyte morphologies, and \textit{vice versa}, that the confined polyelectrolyte morphologies can be used to monitor the change in the relative permittivity of the outside medium.

\section{Methods}
\label{sec:model}
\subsection{Model}
We develop a model system of a linear polyelectrolyte with explicit counterions confined in a spherical cavity. Example physical systems of interest include single-stranded DNAs, RNAs and macromolecules that are highly charged. In these systems, the interior medium of the cavity typically contains water molecules, which are necessary to dissolve the polyelectrolyte and counterions. While the dielectric response of the interior medium is not necessarily uniform, it is reasonable to the first order of approximation that it is in the range of 10--40, \textit{i.e.}, the value at the surface of many proteins.\cite{Li13}  The exterior of the cavity is composed of thick layers of amphiphilic molecules with hydrophilic head groups sticking inward to interact with the essential water molecules, charged monomers and counterions, and hydrophobic groups pointing outward.

\begin{figure}[ht!]
  \centering
  \includegraphics[width=0.3\textwidth, trim=0cm 0cm 0cm 0cm, clip=true]{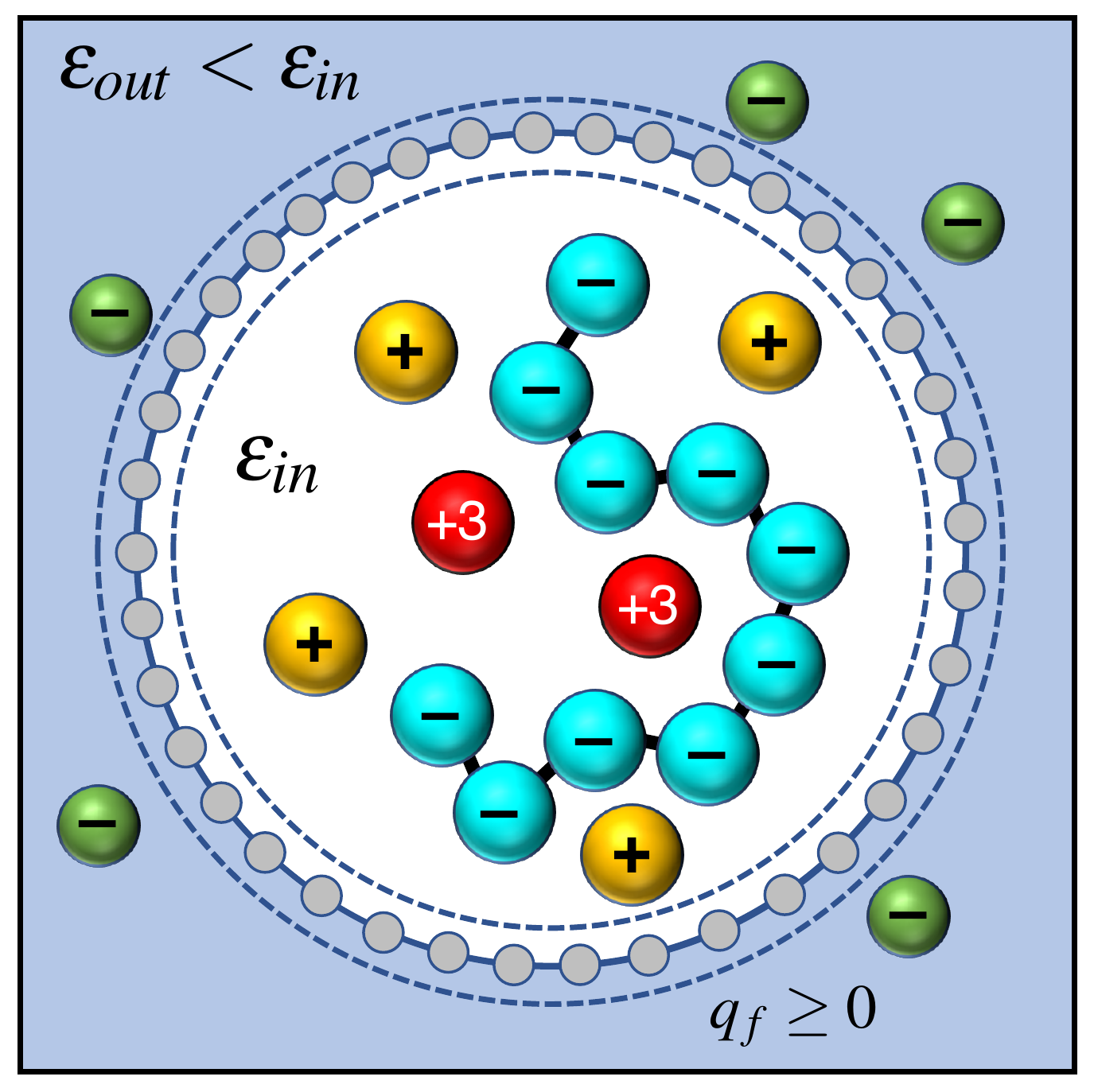}
  \caption{Simulation setup for a polyelectrolyte with monovalent counterions and trivalent counterions confined in a spherical cavity. The media inside and outside the cavity are assumed to be uniform with relative permittivities, $\epsilon_{in}$ and $\epsilon_{out}$, respectively. The dash lines represent the cavity thickness where counterions and charged monomers are excluded.}
  \label{fig:model}
\end{figure}

The coarse-grained model of the confined charged polyelectrolyte system is shown in Fig. \ref{fig:model}. The polymer chain is represented by a bead-spring model \cite{Kremer90} composed of $N = 100$ coarse-grained monomers, each approximately 5-10 \AA \, in diameter. \textcolor{black}{As such, the bead diameter maps roughly to one Kuhn length.} We add $N^{+}_c$ monovalent counterions and $N^{3+}_c$ trivalent counterions to neutralize the total charge of the polymer: $N^{+}_c + 3N^{3+}_c = N$. The spherical interface has radius of $R$, discretized into $M$ vertices (or patches). The polymer beads and counterions interact with the spherical confinement wall \textit{via} the Weeks--Chandler--Andersen (WCA) potential.\cite{WCA} The multivalent counterion fraction is defined as $\phi_m = N^{3+}/N$. We use the same value of $\sigma = 1.0$ for the WCA interaction between all the counterions to represent their hydrodynamic diameters.

For charged interfaces ($q_f > 0$), negatively charged counterions are added in the outer medium to ensure charge neutrality. In our model, we assume that the counterions do not diffuse through the cavity shell, which is relevant for the cases where the shell thickness is sufficiently thick to make the rate of ion transport extremely low, and/or when the Gibbs-Donnan equilibrium of the counterions inside and outside the cavity has been reached. The outer counterion concentration is varied from $0.004\sigma^{-3}-0.02\sigma^{-3}$, corresponding to 20--100 mM. The thickness of the cavity shell, $\delta$, is varied between $1.0-2.0\sigma$ by setting an outer spherical wall with radius $R_{out} = R + \delta$ that interacts with the outer counterions \textit{via} the WCA potential.

The electrostatic interaction strength inside the cavity is characterized by the Bjerrum length, $l_B = q^2\sigma/\epsilon_{in}$, where $q = Ze/\sqrt{4\pi \epsilon_0 \sigma k_BT}$ is the dimensionless charge of the monovalent counterions ($Z = 1$). Unless otherwise stated, the relative permittivity of the medium inside and outside the cavity are $\epsilon_{in} = 40$ and $\epsilon_{out} = 4$, respectively. For $\epsilon_{in} = 40$, the monomer charges $q$ are set so that $l_B/\sigma = 4$, reminiscent of single stranded DNAs and NaPSS. The polyelectrolyte and counterions are equilibrated in the canonical ensemble (constant volume and temperature) using the Nose-Hoover chain thermostat at $k_BT/\epsilon = 1$, where $\epsilon = 1.0$ is the well depth of the WCA potential. The dimensionless time unit is $\tau = \sigma\sqrt{m/\epsilon}$, where $m = 1$ is the bead mass, which is the same for all the particles. The integration timestep is $\Delta t = 0.001\tau$.

To characterize the conformational behavior of the confined polyelectrolyte, we perform umbrella sampling simulations using SSAGES, a software suite for accelerated sampling,\cite{SSAGES} coupled with LAMMPS.\cite{Plimpton95} Essentially, SSAGES employs LAMMPS for sampling the polyelectrolyte conformation in the canonical ensemble, and at every time step applies the biasing forces to the charged monomers based on the difference between the instantaneous value of the collective variable, $R^2_g$ in this case, and the constrained value set within each window, $R^2_{g0}$. The biasing potential applied to the monomers at every time step is given by $U^{(b)} = \dfrac{1}{2}k(R^2_g - R^2_{g0})^2$, where $k$ is the biasing constant, which gives the biasing force on individual monomers: $\mathbf{f}^{(b)}_i = -\nabla_{\mathbf{r}_i} U^{(b)}$. We tested with $k = 50$, 75 and 100 (in unit of $\epsilon/\sigma^4$), and chose $k = 75$ for all of the simulations reported to ensure that the histograms of $R^2_g$ obtained from the adjacent windows highly overlap.

We use 32--48 windows for umbrella sampling, each corresponding to a constrained value of the polyelectrolyte squared radius of gyration, $R^2_{g0}$. The polarization solver in use is the \textcolor{black}{induced} charge computation (ICC$^*$) method\cite{Tyagi10} implemented in LAMMPS by our previous work.\cite{TrungCPC19} The biased histograms $P_b(R^2_g)$ obtained from individual windows are then combined together using the weighted histogram analysis method \cite{WHAM} to yield the free energy profile. Note that all the free energy curves are shifted to zero at their minimum in the WHAM calculations. In the present study, we have performed more than 500 replica simulations in total, each equilibrated for 5000-10000$\tau$. By fitting the autocorrelation function of the system potential energy with $\exp(-t/\tau_r)$, we found that the relaxation time $\tau_r$ is found to be on the order of the time unit $\tau$ for the simulated systems. There are more than 1000 simulations (as replicas and as independent runs) in total performed in the present study.

\section{Acknowledgments}
The research was supported by the Sherman Fairchild Foundation and by the Center for Computation and Theory of Soft Materials. \textcolor{black}{We thank Michael Engel for sharing the code used to generate the diffraction patterns.}
%in Figure \ref{fig:phasediagram}.

\bibliographystyle{unsrt}
\bibliography{references}

\clearpage

\renewcommand{\thefigure}{S\arabic{figure}}
\setcounter{figure}{0}

\section{Supporting Information}
\subsection{Supporting Movie}
Movie S1: Four-fold conformation of the 100-mer polyelectrolyte with the counterions filled in the quarters. The cavity radius is $R = 6\sigma$. The surface charge density is $q_f\sigma^2/q = 0.2$. Trivalent counterion fraction is $\phi_m = 0.3$. The monomers are in cyan, monovalent counterions in orange, trivalent counterions in red, outer counterions in green; and small white spheres are interface beads. For clarity we only show the upper half of the interface particles (small white spheres) and outside counterions (green spheres) to better reveal the polyelectrolyte conformations and counterions inside the cavity.

\subsection{Supporting Figures}

\begin{figure*}[ht!]
  \centering
  \includegraphics[width=1.0\textwidth, trim=0cm 0cm 0cm 0cm, clip=true]{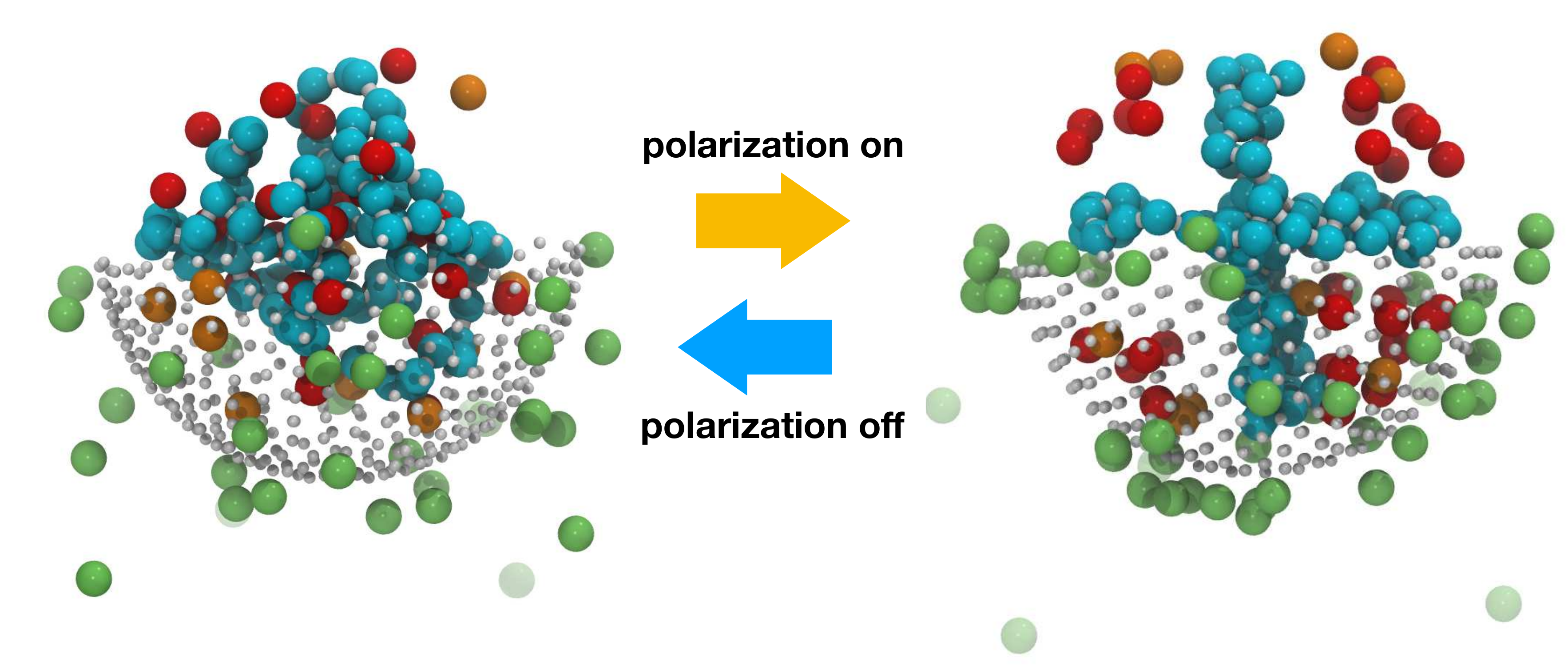}
  \caption{Reversible transformations between amorphous to four-fold morphologies when polarization effects are turned on and off. The cavity radius is $R = 6\sigma$ and trivalent counterion fraction $\phi_m = 0.3$. $\sigma$ is the length scale of the exclude volume interaction between the monomers and the counterions (i.e. the Weeks--Chandler--Andersen potential). The monomers are in cyan, monovalent counterions in orange, trivalent counterions in red, outer counterions in green; and small white spheres are interface beads. Here and in the snapshots that follows, for clarity we only show the upper half of the interface particles (small white spheres) and outside counterions (green spheres) to better reveal the polyelectrolyte conformations and counterions inside the cavity.}
  \label{fig:transforms}
\end{figure*}

\begin{figure*}[ht!]
  \centering
  \includegraphics[width=0.7\textwidth, trim=0cm 0cm 0cm 0cm, clip=true]{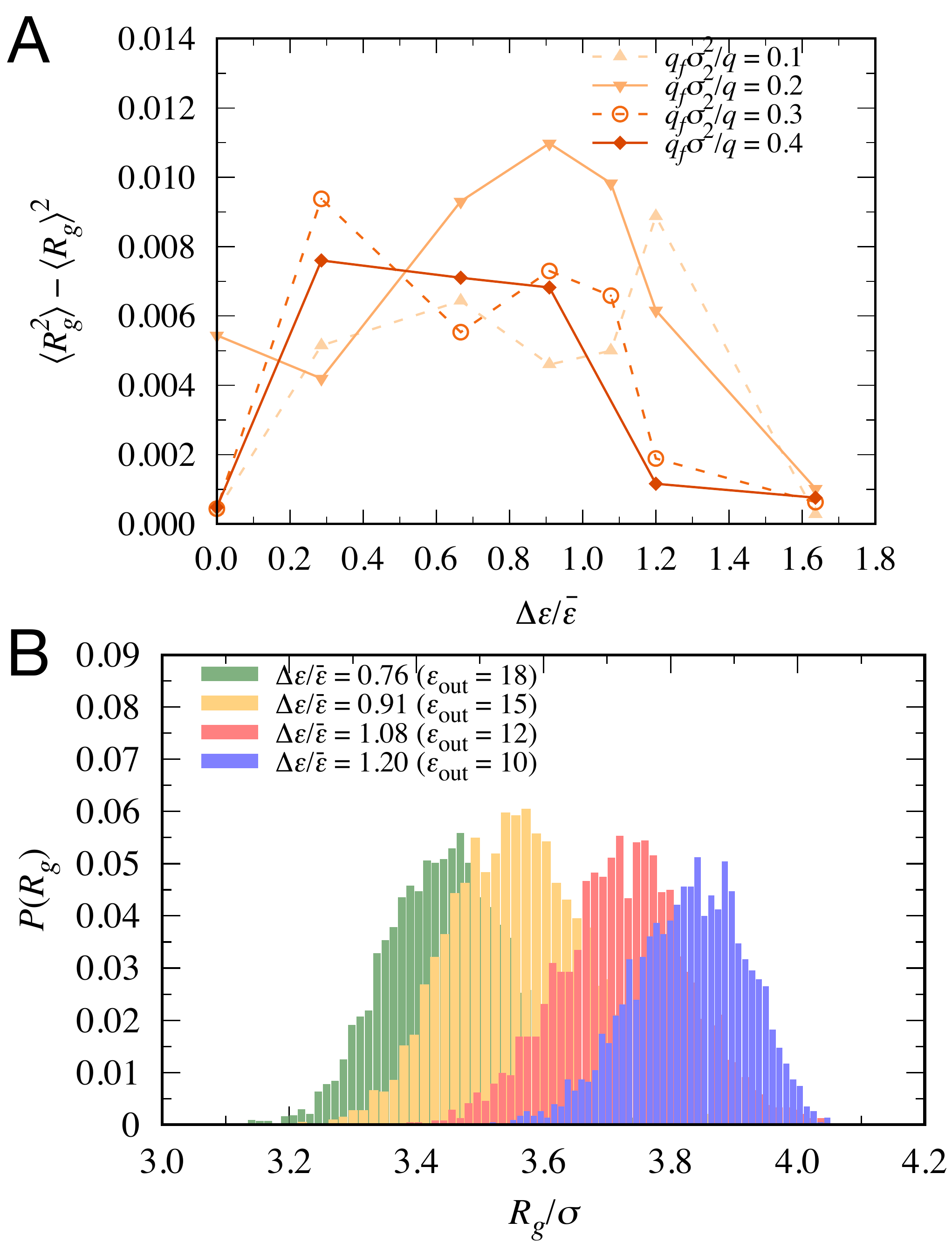}
  \caption{A) Fluctuations in the polyeletrolyte radius of gyration as a function of dilectric mismatch for different surface charge densities $q_f\sigma^2/q$, where $q$ is the monomer charge, and $\sigma$ is the distance unit of the exclude volume interaction between the monomers (i.e. the Weeks-Chandler-Andersen potential). B) Polyelectrolyte radius of gyration for different values of outside the relative permittivity $\epsilon_{out}$ and the corresponding dielectric mismatch $\Delta \epsilon/\bar{\epsilon}$ from unbiased simulations. The surface charge density is $q_f\sigma^2/q = 0.2$. The cavity radius is $R = 6\sigma$ and trivalent counterion fraction $\phi_m = 0.3$.}
  \label{fig:Rgvariance}
\end{figure*}

\begin{figure*}[ht!]
  \centering
  \includegraphics[width=1.0\textwidth, trim=0cm 0cm 0cm 0cm, clip=true]{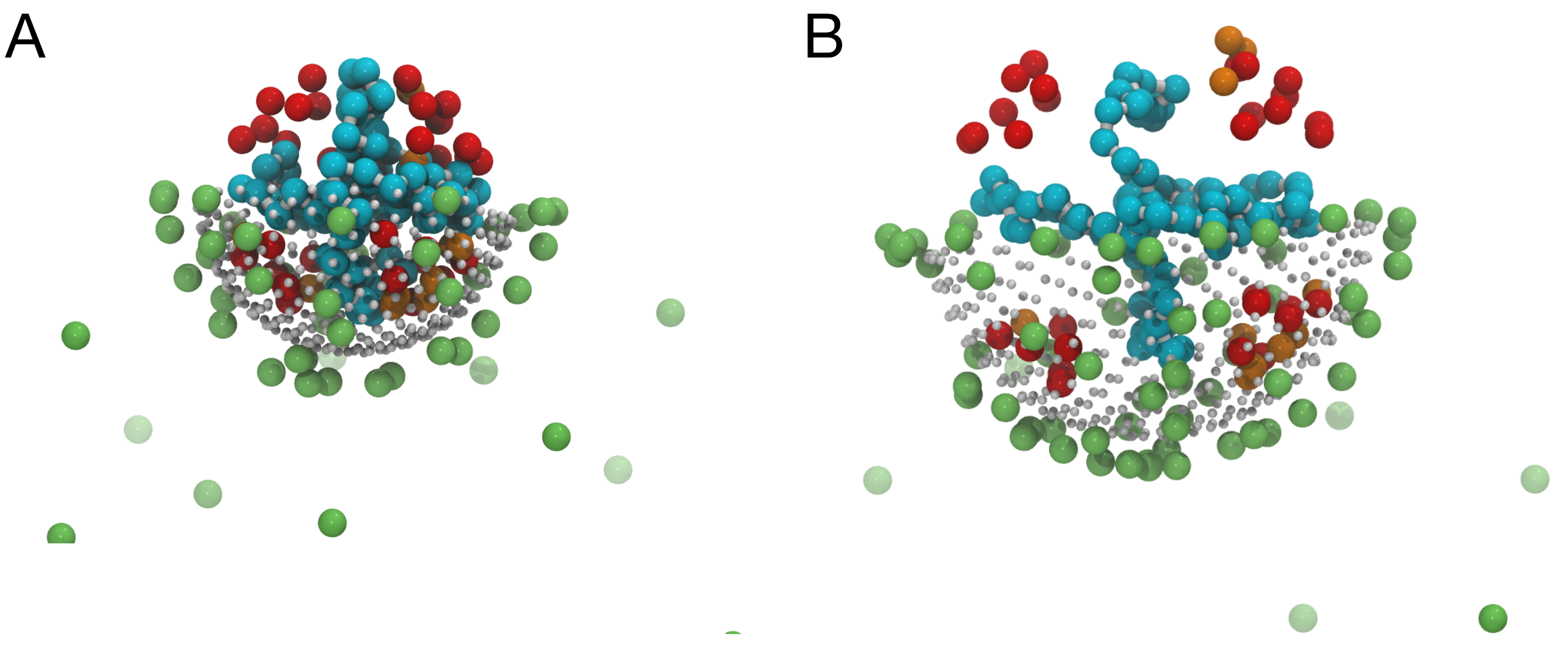}
  \caption{Representative snapshots of the equilibrium conformations of the polyelectrolyte for different cavity radii:  A) $R = 5\sigma$ and B) $R = 7\sigma$. The surface charge density is $q_f\sigma^2/q = 0.2$. Trivalent counterion fraction is $\phi_m = 0.3$.}
  \label{fig:radii}
\end{figure*}

\begin{figure*}[ht!]
  \centering
  \includegraphics[width=0.9\textwidth, trim=0cm 0cm 0cm 0cm, clip=true]{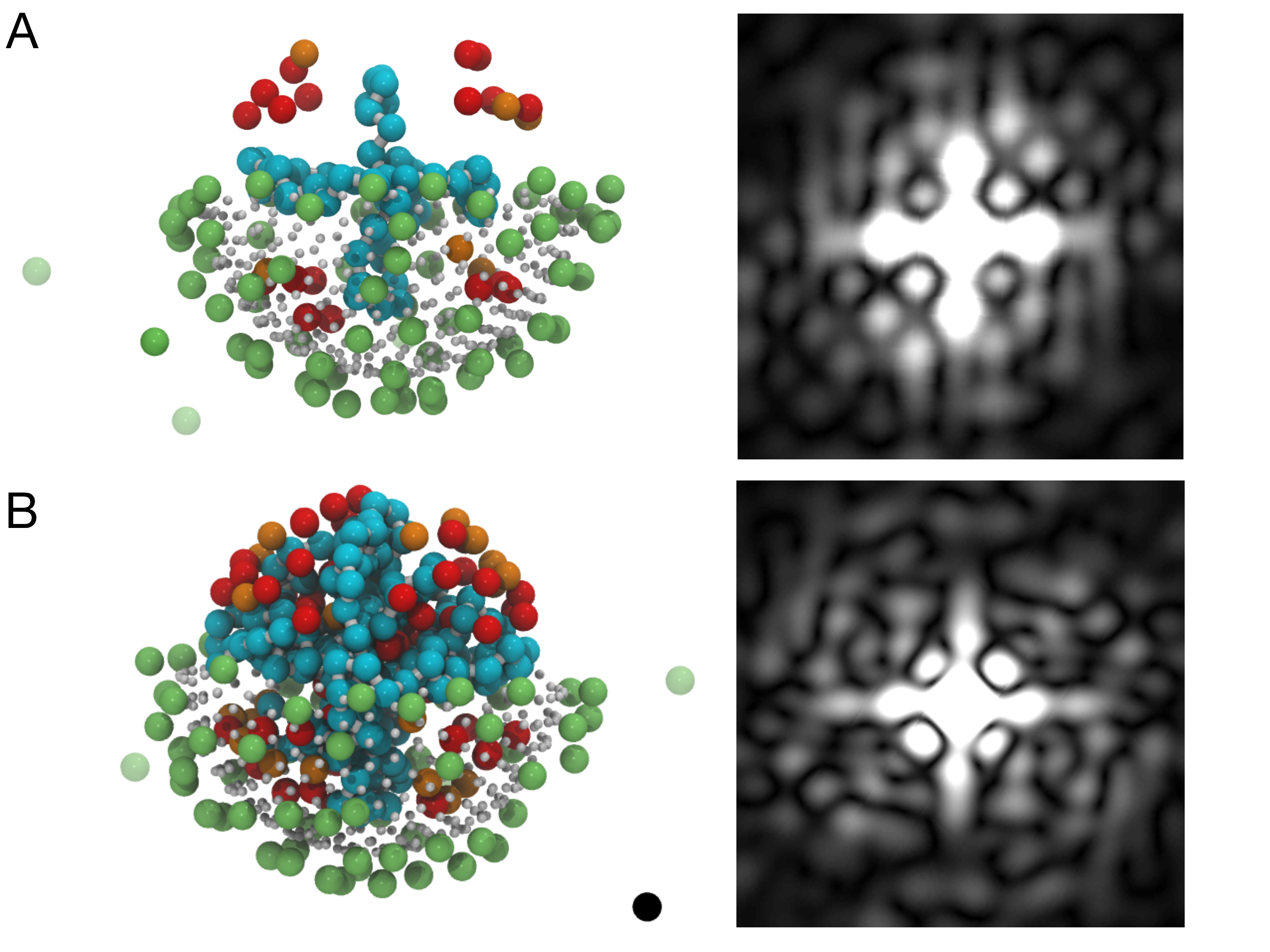}
  \caption{Representative snapshots of the equilibrium conformations of the polyelectrolyte for different chain lenghths: A) $N = 60$; B) $N = 200$. Insets are the diffraction pattern taken along the four-fold symmetry axis. The cavity radius is $R = 6\sigma$. Trivalent counterion fraction is $\phi_m = 0.3$.}
  \label{fig:chainlengths}
\end{figure*}

\begin{figure*}[ht!]
  \centering
  \includegraphics[width=1.0\textwidth, trim=0cm 0cm 0cm 0cm, clip=true]{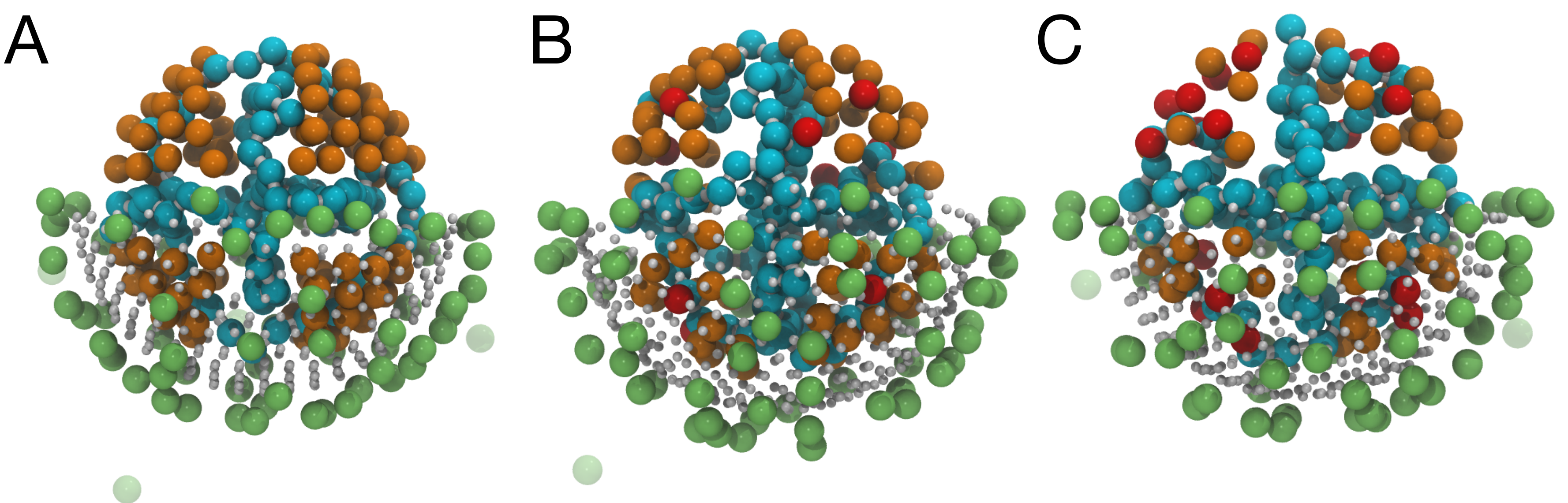}
  \caption{Representative snapshots of the conformations of the polyelectrolyte for different trivalent counterion fractions:
A) $\phi_m = 0$; B) $\phi_m = 0.1$ and C) $\phi_m = 0.2$. The chain length is $N = 100$. The surface charge density is $q_f\sigma^2/q = 0.2$.
The cavity radius is $R = 6\sigma$.}
  \label{fig:ftri}
\end{figure*}

\begin{figure*}[ht!]
  \centering
  \includegraphics[width=1.0\textwidth, trim=0cm 0cm 0cm 0cm, clip=true]{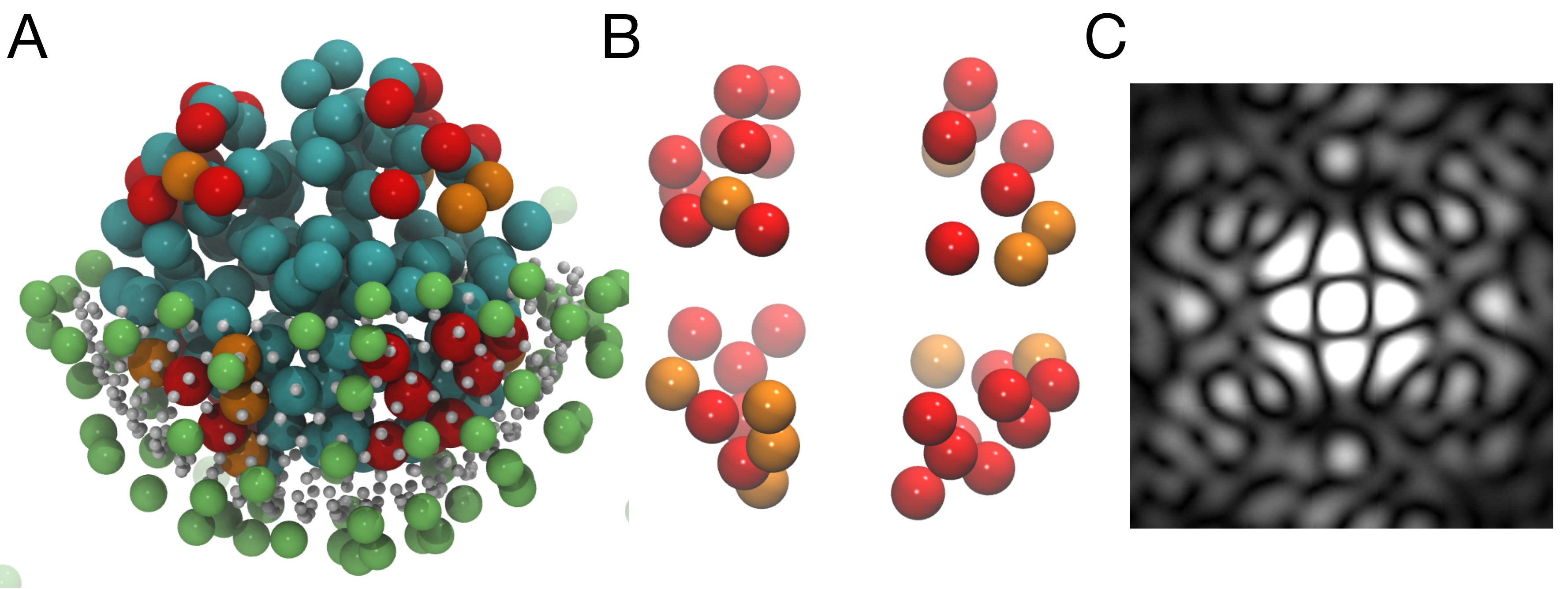}
  \caption{A) Representative equilibrated configuration of the monomers when all the bonds are removed. B) Only the counterions are visualized to show the simple cubic arrangement, as shown in the diffraction pattern in C). The dimensionless surface charge density is $q_f\sigma^2/q = 0.2$, and trivalent counterion fraction $\phi_m = 0.3$.}
  \label{fig:nobonds}
\end{figure*}

\begin{figure*}[ht!]
  \centering
  \includegraphics[width=0.4\textwidth, trim=0cm 0cm 0cm 0cm, clip=true]{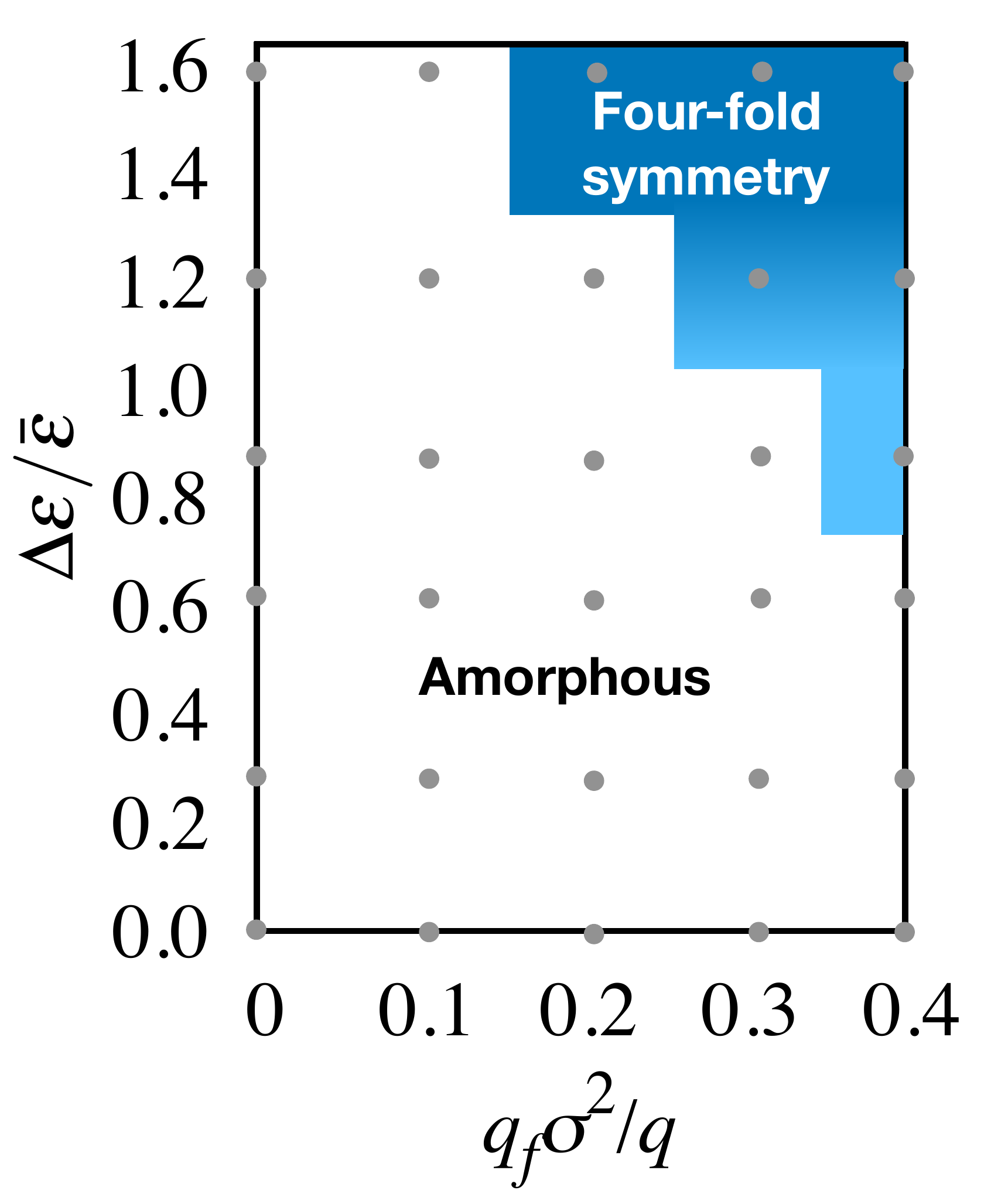}
  \caption{Phase diagram of equilibrium morphologies of the confined polyelectrolyte as function of the surface charge density ($q_f\sigma^2/q$) and dielectric mismatch ($\Delta \epsilon/\bar{\epsilon}$). Dots represent the state points where simulations are performed. The chain length is $N$ = 100. The cavity radius is $R = 6\sigma$ and trivalent counterion fraction $\phi_m = 0.1$.}
  \label{fig:phasediagram01}
\end{figure*}

\begin{figure*}[ht!]
  \centering
  \includegraphics[width=1.0\textwidth, trim=0cm 0cm 0cm 0cm, clip=true]{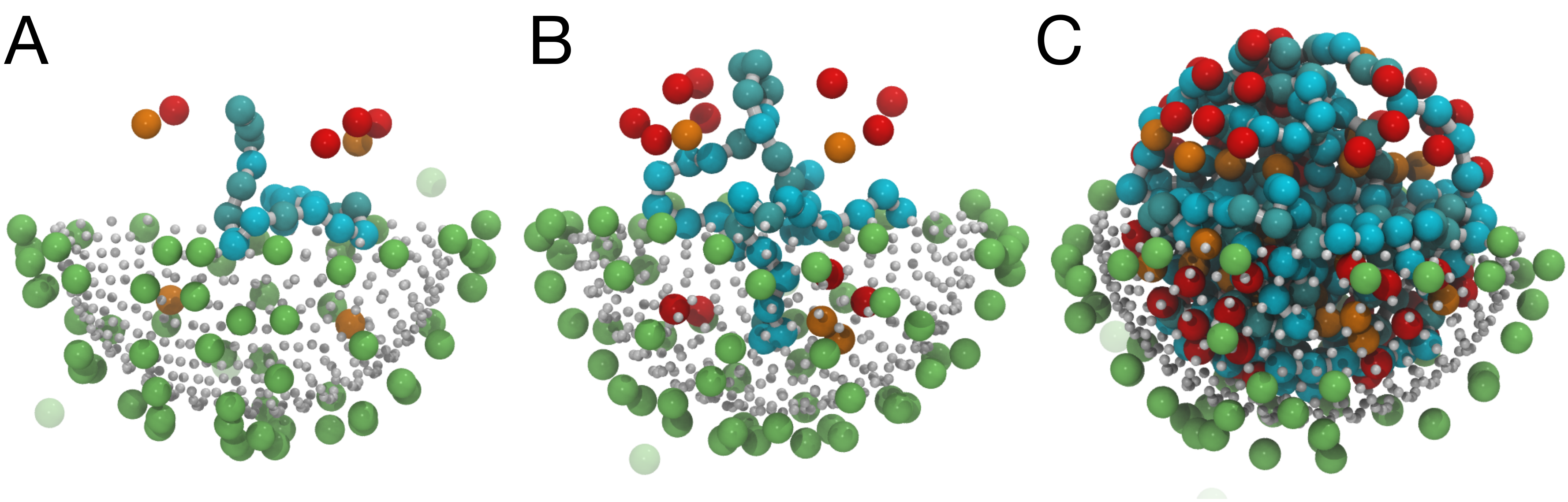}
  \caption{Representative snapshots of the conformations of the polyelectrolyte for extreme chain lengths: A) $N = 20$; B) $N = 40$ and C) $N = 250$.
The cavity radius is $R = 6\sigma$. Trivalent counterion fraction is $\phi_m = 0.3$.}
  \label{fig:chainlengths2}
\end{figure*}

\begin{figure*}[ht!]
  \centering
  \includegraphics[width=0.8\textwidth, trim=0cm 0cm 0cm 0cm, clip=true]{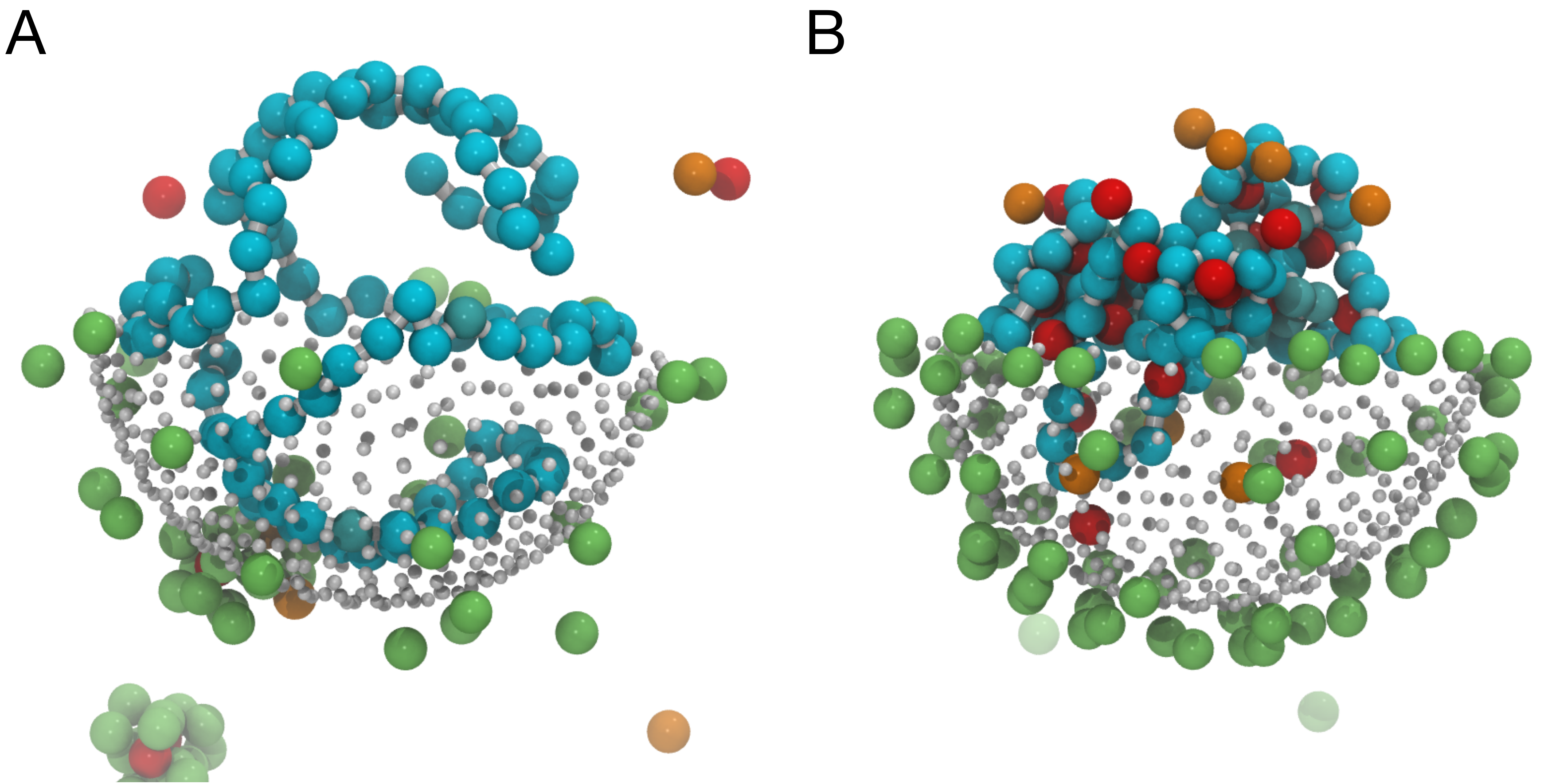}
  \caption{Four-fold symmetry conformations do not emerge A) when only the polyelectrolyte is confined inside the cavity while all the counterions are located outside; and when B) the relative permittivity inside the cavity is reduced to $\epsilon_{in} = 10$, while $\epsilon_{in} = 4$.
The chain length is $N = 100$, the cavity radius is $R = 6\sigma$ and trivalent counterion fraction is $\phi_m = 0.3$.}
  \label{fig:amorphous}
\end{figure*}

\begin{figure*}[ht!]
  \centering
  \includegraphics[width=0.7\textwidth, trim=0cm 0cm 0cm 0cm, clip=true]{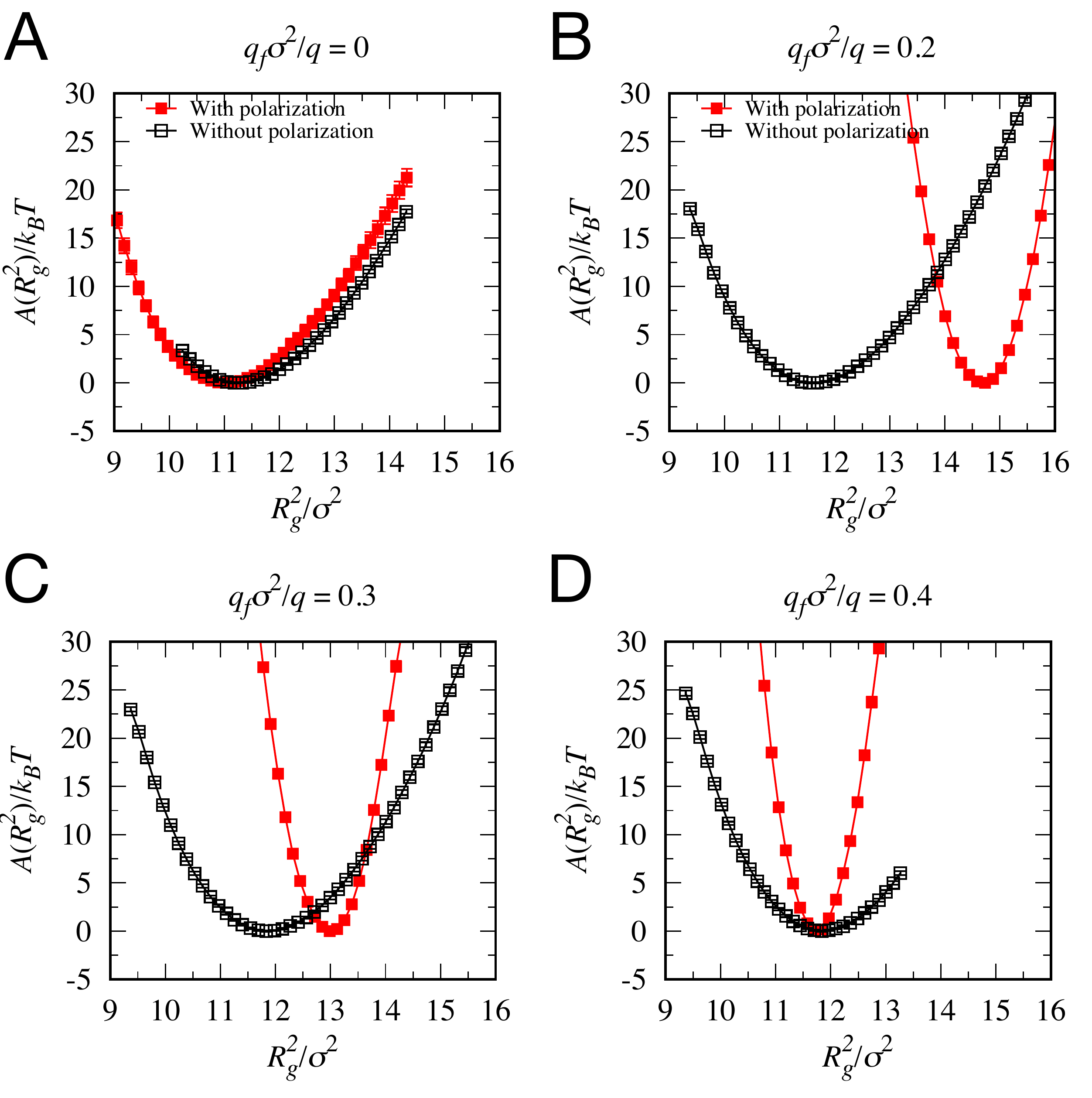}
  \caption{Free energy profiles of the polyelectrolyte (squared) radius of gyration for different dimensionless surface charge densities $q_f$. ``With polarization'' corresponds to $\epsilon_{in} = 40$ and $\epsilon_{out} = 4$; ``Without polarization'' corresponds to $\epsilon_{in} = \epsilon_{out} = 40$, or when polarization effects are excluded. Trivalent counterion fraction is $\phi_m = 0.3$. Error bars are the statistical errors obtained from bootstrapping analysis using WHAM.}
  \label{fig:pmfqsurfs}
\end{figure*}

\begin{figure*}[ht!]
  \centering
  \includegraphics[width=1.0\textwidth, trim=0cm 0cm 0cm 0cm, clip=true]{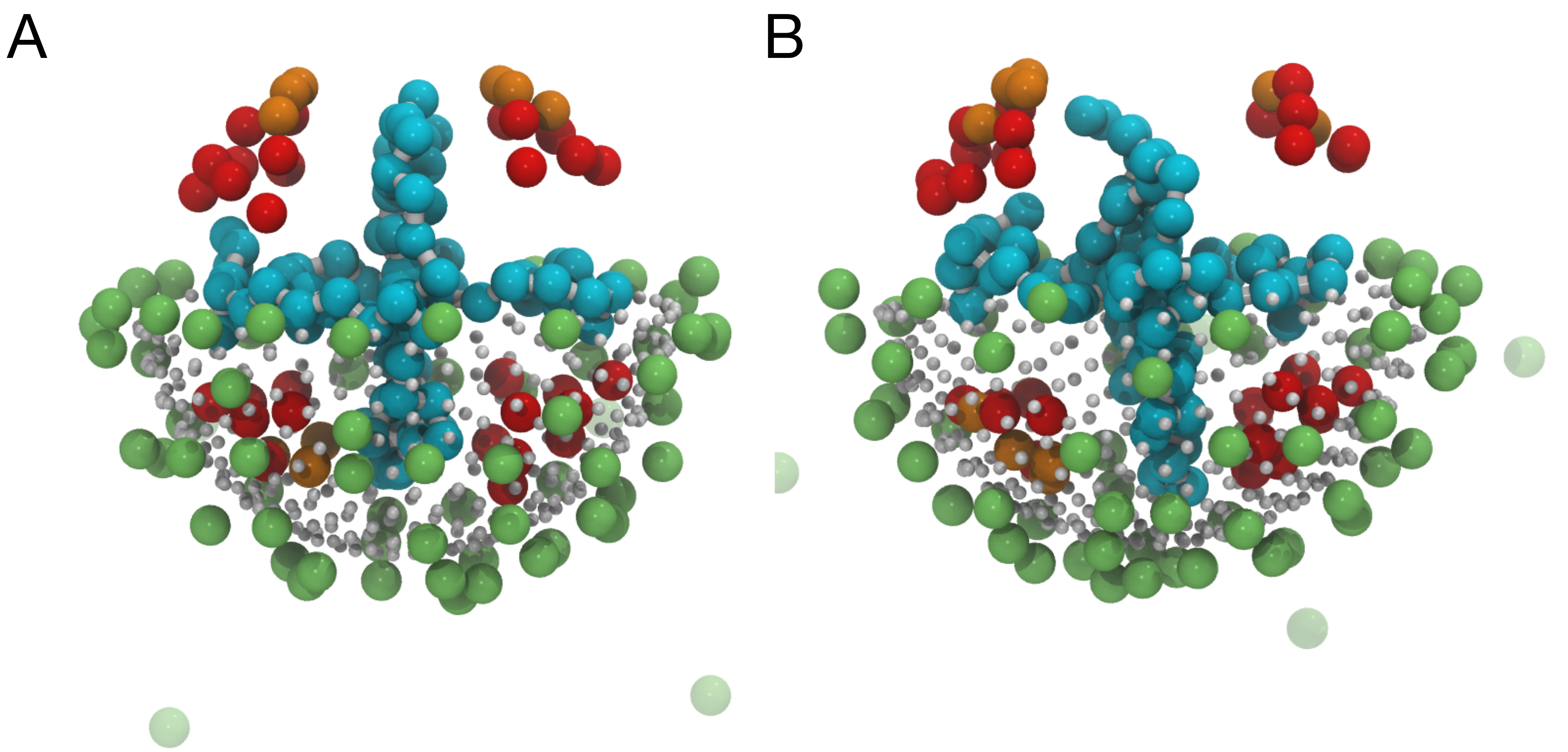}
  \caption{Representative snapshots of the equilibrium conformations of the polyelectrolyte for different dimensionless surface charge densities: A) $q_f\sigma^2/q = 0.3$; B) $q_f\sigma^2/q = 0.4$, where $q$ is the monomer charge and $\sigma$ is the distance unit of the exclude volume interaction between the monomers (i.e. the Weeks-Chandler-Andersen potential). Trivalent counterion fraction is $\phi_m = 0.3$.}
  \label{fig:qsurfs}
\end{figure*}

\begin{figure*}[ht!]
  \centering
  \includegraphics[width=0.8\textwidth, trim=0cm 0cm 0cm 0cm, clip=true]{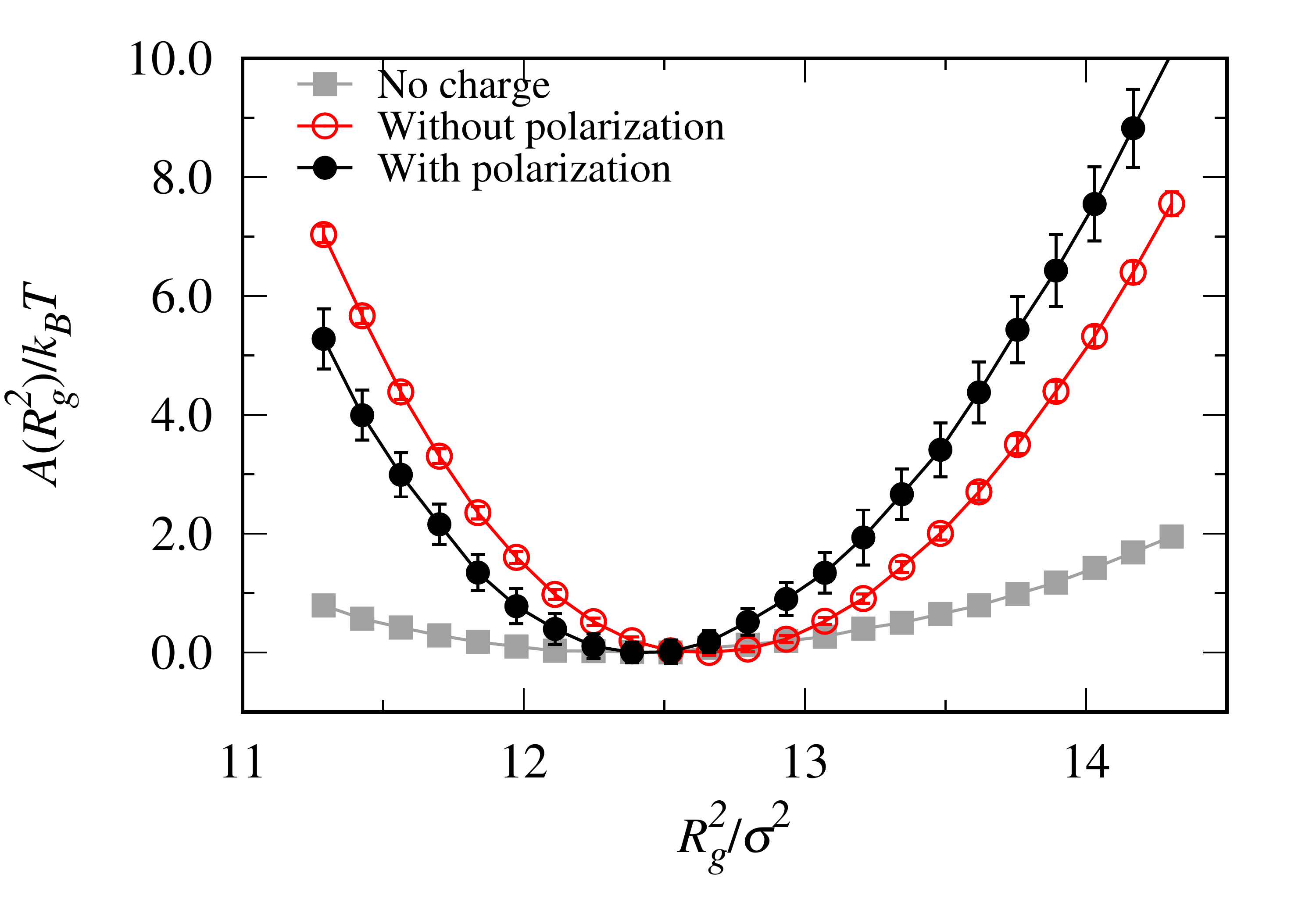}
  \caption{Representative equilibrated configuration of the monomers when the cavity surface is charge neutral $q_f = 0$. Error bars are the statistical errors obtained from bootstrapping analysis using WHAM. Trivalent counterion fraction is $\phi_m = 0.1$.}
  \label{fig:neutral}
\end{figure*}

\begin{figure*}[ht!]
  \centering
  \includegraphics[width=0.7\textwidth, trim=0cm 0cm 0cm 0cm, clip=true]{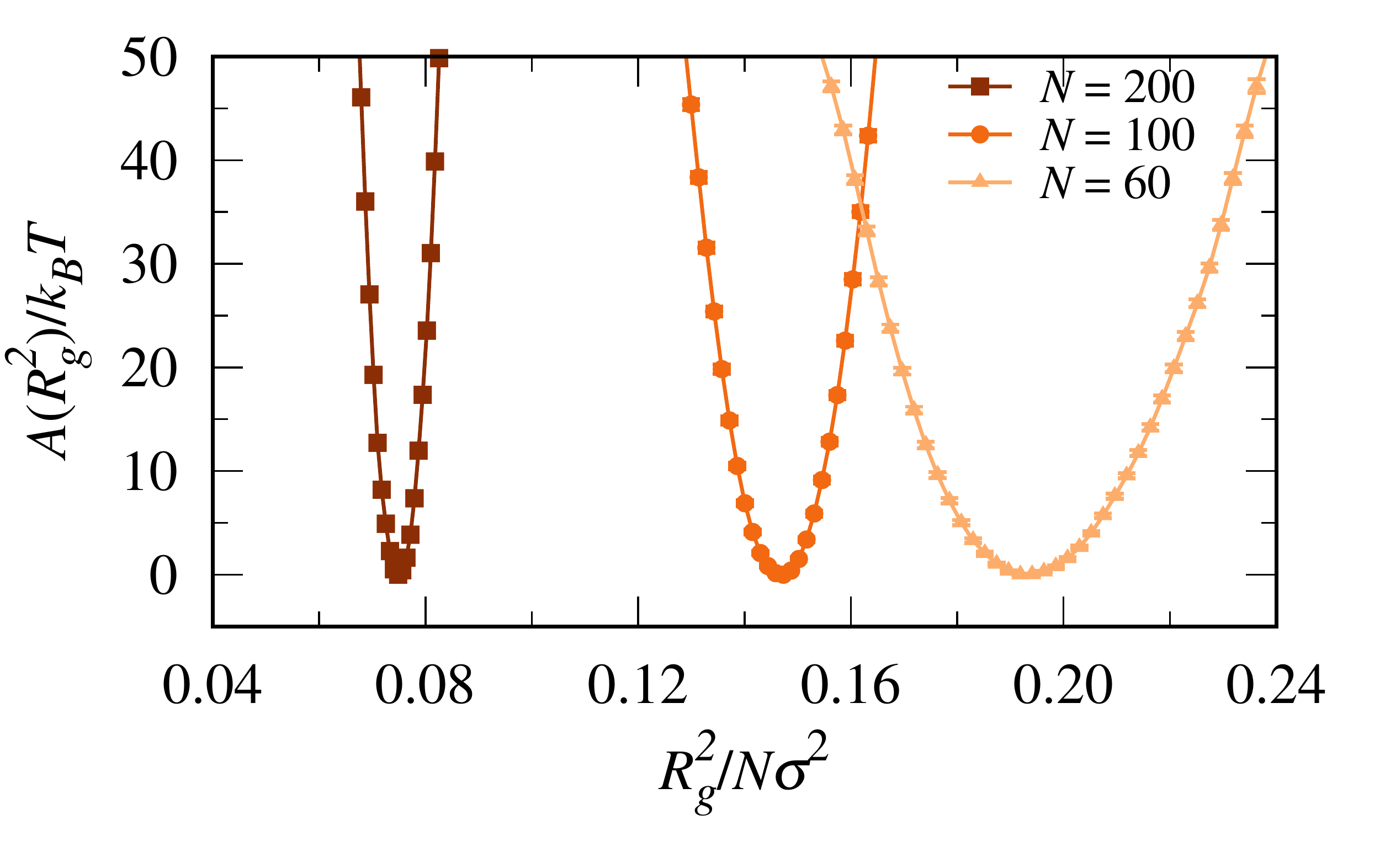}
  \caption{Free energy profiles of the polyelectrolyte (squared) radius of gyration for different chain lengths $N$ for the same cavity radius $R = 6\sigma$.
Trivalent counterion fraction is $\phi_m = 0.3$.}
  \label{fig:pmfchainlengths}
\end{figure*}

%\begin{figure*}[ht!]
%  \centering
%  \includegraphics[width=0.75\textwidth, trim=0cm 0cm 0cm 0cm, clip=true]{SI-histograms-Rg-ein.pdf}
%  \caption{Polyelectrolyte radius of gyration for different values of outside the relative permittivity $\epsilon_{out}$ and the corresponding dielectric mismatch $\Delta \epsilon/\bar{\epsilon}$ from %unbiased simulations. The surface charge density is $q_f\sigma^2/q = 0.2$, the cavity radius is $R = 6\sigma$ and trivalent counterion fraction $\phi_m = 0.3$.}
%  \label{fig:histRgein}
%\end{figure*}

\begin{figure*}[ht!]
  \centering
  \includegraphics[width=0.8\textwidth, trim=0cm 0cm 0cm 0cm, clip=true]{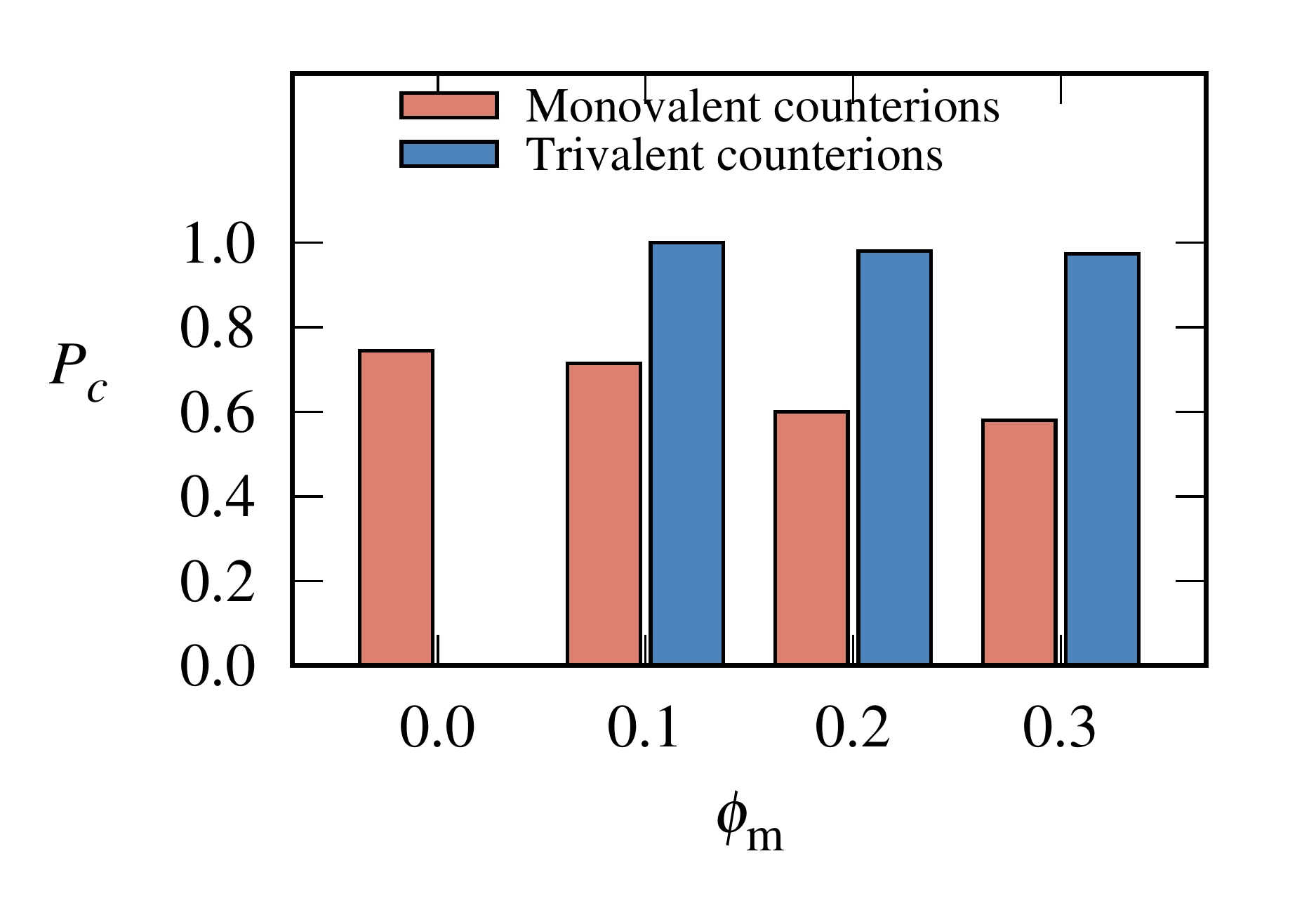}
  \caption{Probability of observing condensed counterions as a function of trivalent counterion fraction, $\phi_m$. A counterion is considered condensed if it is within a cutoff distance of $1.2\sigma$ from a monomer of the polyelectrolyte. The cavity radius is $R = 6\sigma$.  The cavity surface is charge neutral $q_f = 0$.}
  \label{fig:probability}
\end{figure*}

\end{document}